\documentclass{emulateapj}
\usepackage{graphicx}
\usepackage{times}
\usepackage{lscape} 

\received{28 April 2006}
\accepted{31 May 2006}

\shortauthors{SCHMIDT, CONNOLLY, \& HOPKINS}
\shorttitle{DRAGONS}

\begin{document}

\title{THE DRAGONS SURVEY: A SEARCH FOR HIGH REDSHIFT RADIO GALAXIES AND HEAVILY OBSCURED AGNS\footnotemark[1] }

\author{S. J. SCHMIDT$^{1}$,A. J. CONNOLLY$^{1}$,A. M. HOPKINS$^{1,2}$
}

\affil{
\begin{enumerate}
\item Dept.\ of Physics and Astronomy, University
 of Pittsburgh, 3941 O'Hara Street, Pittsburgh, PA 15260
\item School of Physics, University of Sydney, Bldg A28, NSW 2006, Australia 
\end{enumerate}
}

\begin{abstract}
We present the first results from the Distant Radio Galaxies Optically
Non-detected in the SDSS ({\em DRaGONS}) Survey. Using a novel selection
technique for identifying high redshift radio galaxy (HzRG) candidates, a large
sample is compiled using bright ($S_{\rm 1.4GHz} >\,100$mJy) radio sources
from the Faint Images of the Radio Sky at Twenty centimeters ({\em FIRST})
survey having no optical counterpart in the Sloan Digital Sky Survey
({\em SDSS}). Near-infrared (NIR) $K$-band imaging with the {\em FLAMINGOS}
instrument on the 4-meter telescope at Kitt Peak for 96 such candidates
allows preliminary identification of HzRG candidates through the well-known
$K-z$ relation, and these objects will subsequently be observed
spectroscopically to confirm their redshifts. Of the initial candidates,
we identify 70 with magnitudes brighter than $K\,\approx\,19.5$, and compute
limiting magnitudes for the remainder.  Assigning redshifts based on a
linear fit to the $K-z$ Hubble diagram gives a mean redshift for
our sample of $z=2.5$ and a median redshift of $z=2.0$, showing that this
method should be very efficient at identifying a large number of HzRGs.
This selection is also sensitive to a previously unseen population of anomalously red radio galaxies ($r-K\,>\,6.5-7$), 
which may indicate significant obscuration at moderate redshifts.  These obscured objects can be used to test the completeness 
of QSO surveys to the effects of reddenning.  More than ten percent of our sample falls into this category, which may 
represent a sizable radio loud population missing from current optically selected AGN samples. 
We additionally identify 479 bright Extremely Red Objects (EROs) in the
fields surrounding our HzRG candidates, to a magnitude of $K=17.5$, within
a non-contiguous area of 2.38 square degrees.  This constitutes a small
overdensity of EROs surrounding the radio galaxy candidates over random
fields, and we see possible evidence for a physical association of
the EROs with the radio galaxies. However, examining the clustering of {\em all}
$K\le19.0$ galaxies around the radio targets reveals no evidence
of a global galaxy excess, strengthening our conclusion that the EROs trace an overdensity not 
evident in the overall galaxy population.
\end{abstract}


\section{Introduction}
\label{int}
\footnotetext[1]{Based on observations collected at Kitt Peak National Observatory, National Optical Astronomy Observatory, which is operated by the Association of Universities for Research in Astronomy, Inc. (AURA) under cooperative agreement with the National Science Foundation.}
Radio galaxies have long been used to probe the epoch of galaxy formation
\citep[]{Lil:84,Gra:96,Blu:98,Ste:99,Jar:01}. Luminous radio galaxies
are known to be highly biased relative to the underlying dark matter,
residing in the most overdense regions of the Universe.  Under the
standard Cold Dark Matter hierarchy for galaxy formation these galaxies
should be the first systems to collapse and, therefore, the site of
some of the most evolved stellar populations.  As the likely hosts of
the earliest star formation, it may also be possible to probe the epoch
of reionization \citep[]{Bark:05}.  Isolating a sample of radio galaxies
at high redshift would allow us to probe the physical processes that
drive the formation and evolution of structure and the timescales that
govern star formation in the early universe, as well as to distinguish
between hierarchical and ``down-sizing" \citep[]{Cow:96} formation
scenarios \citep[e.g.,][]{Roc:04,Del:05}, and also the role of feedback
on the local radio galaxy environment \citep[]{Var:05, Del:05}.

While radio surveys can address many fundamental questions in cosmology
and galaxy formation the numbers of radio galaxies identified at
high redshift remain small in comparison to low redshift surveys
\citep[]{Bra:05,Mag:04}.  Without large, statistically complete and
homogeneously selected samples we cannot hope to constrain hierarchical
galaxy formation models without the concern that sample variance might
bias our analyses.  For example, at redshifts $z>3$ extensive optical
and near-infrared (NIR) campaigns  have yielded less than one hundred galaxies
\citep[]{VanB:98,Deb:01,Var:05}.  The reason for the paucity of these
samples comes from the necessity of surveying large volumes to identify
the most massive systems.  Given the broad redshift distribution of
radio galaxies \citep[e.g.,][]{DP:90}, large numbers of radio targets must be
observed in order to extract the high redshift component.

In this paper we describe a novel selection technique that utilizes
existing optical and radio surveys to overcome these challenges in
order to isolate high redshift candidates for follow up in the NIR.
We show that this approach gives more than a factor of ten
increase in efficiency over blind radio selected surveys.  The resulting
combination of NIR data with existing optical measurements
from SDSS allows us to study the environment of the high redshift radio
galaxies.  Early in the development of infrared observations it was
noted that EROs are often found in the vicinity of high redshift objects
\citep[]{Mcc:92,Gra:94,Dey:95}.  More recently, targeted searches have
been undertaken to search for clustering of galaxies \citep[]{Hal:01}
and extremely red objects (EROs) around high redshift quasars and radio
galaxies \citep[][and references therein]{Cim:00,Wold:03,Zhe:05}. The
focus of these investigations has been whether the increased surface
density of EROs is physically associated with the radio galaxy or quasar,
or whether it lies in the foreground, possibly as a cluster that may
have gravitational lensing effects on the target.

Also of interest in a near-infrared radio survey is the number of obscured quasars.  It is well know that optical selection is sensitive to a wide range of effects that bias samples against the detection of heavily obscured galaxy populations.  Indeed, heavily obscured quasars may account for a significant fraction of the total population \citep[]{Web:95,Whi:03,Gli:04}.  Our selection criteria result in the detection of a population of bright ($K\,<\,17.5$), red ($r-K\,>\,6.5$) objects that may fall in to this category.    

The structure of this first in a series of papers is as follows:  The target selection procedure is presented in \S\,\ref{target}.  A description of the observations, astrometry and photometry is given in \S\,\ref{obs}.  The results are presented and discussed in \S\,\ref{res}, and a summary and discussion of future work is presented in \S\,\ref{summ}.

SDSS magnitudes are in the AB system, while we will use Vega magnitudes for all $K$-band measurements for easy comparison with the high redshift radio galaxy literature.
A flat Lambda cosmology is assumed throughout, as indicated by
numerous recent measurements, with $H_0=70\,$km\,s$^{-1}$\,Mpc$^{-1}$,
$\Omega_M=0.3$, $\Omega_{\Lambda}=0.7$ \citep[e.g.,][]{Spe:03}.

\section{Target Selection}
\label{target}

\begin{figure}
\begin{center}
\includegraphics[width=235pt]{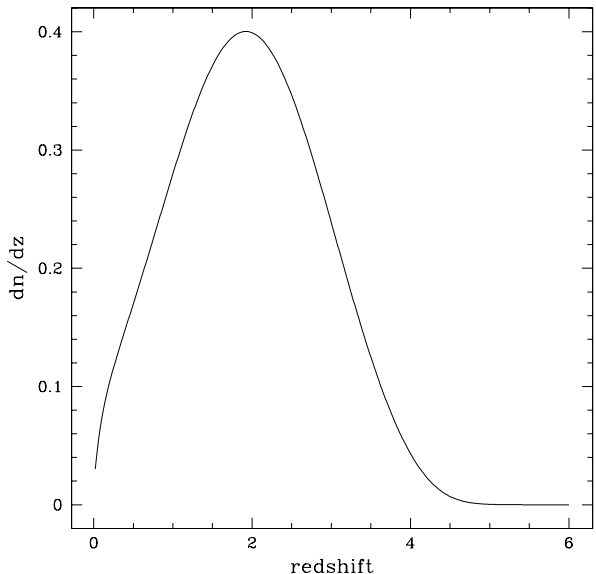}
\caption{$dN/dz$ based on the \citet{DP:90} model for
$S_{\rm 1.4GHz}\,>100\,$mJy radio sources. This shows the broad redshift
distribution expected for bright radio sources.
The y-axis scale is in arbitrary units.
 \label{dndz}}
\end{center}
\end{figure}

The main obstacle to identifying high redshift radio galaxies is screening
out the low redshift foreground.  Figure~\ref{dndz} shows a $dN/dz$
distribution for bright ($S_{\rm 1.4GHz}>100\,$mJy) radio sources based
on model radio luminosity functions and assuming pure luminosity evolution
\citep[]{Row:93,DP:90}. With a broad redshift peak at $z \sim 2$ extracting
only the high-$z$ galaxies through blind spectroscopic follow up of radio
surveys is inefficient.  Our goal is to eliminate the low redshift
contamination through the inclusion of multi-wavelength information.
We begin with the well known $K-z$ Hubble relation for radio galaxies
\citep[]{Lil:84,VanB:98,Jar:01}, which shows the strong correlation of
$K$-band apparent magnitude and redshift.  If we assume that this relation
holds for all bright radio galaxies, then we can use model galaxy colors
to predict the optical properties of these galaxies as they evolve.

\begin{figure*}
\epsscale{1.0}
\plotone{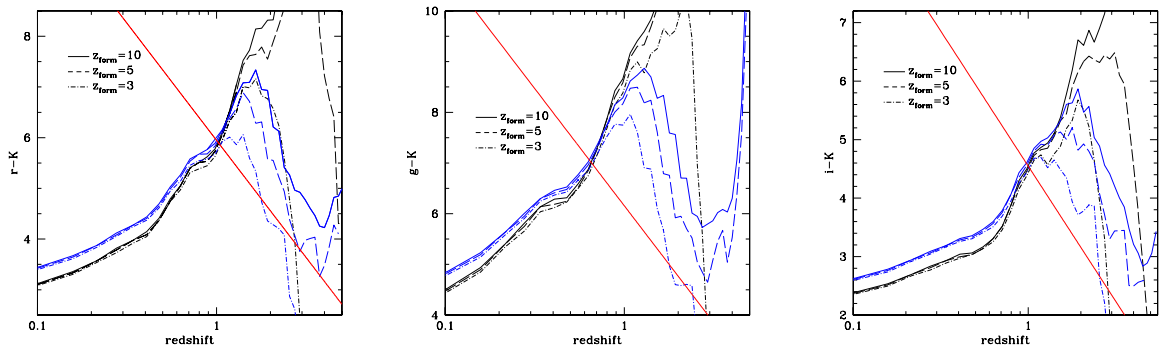}
\caption{Color-redshift diagrams for two PEGASE models at three
formation redshifts.  The upper tracks represent a solar metallicity
instantaneous burst model with no obscuration, while the lower tracks
show a zero initial metallicity exponentially declining star formation
with $\tau=0.5\,$Gyr and ``spheroid" extinction assumed.  The diagonal
lines represent $5\,\sigma$ limits in SDSS $g$, $r$, and $i$ bands.
Objects above the diagonal line will not be detected in the individual
SDSS band.
\label{simplerk}}
\end{figure*}

Figure~\ref{simplerk} shows the $r-K$, $g-K$, and $i-K$ color-redshift
diagrams for two sets of galaxy models generated with the spectral
synthesis code {\em PEGASE} \citep[]{FRV:97}.  The upper curves are color
tracks for an instantaneous burst model at solar metallicity and assuming
no extinction for formation redshifts of $z_{\rm form} = $ 10, 5, and 3, while
the lower curves are for zero initial metallicity with an exponentially
declining star-formation model with an $e$-folding time of $\tau=0.5\,$Gyr
and PEGASE ``spheroid'' type obscuration at these same formation redshifts.
We choose solar metallicity for the instantaneous burst model to match
the $r-K$ color of an elliptical galaxy at low redshift, as the PEGASE models
do not update the metallicity in an instantaneous burst. Metallicity does
evolve in the models of ongoing star formation, hence the assumption of
zero initial metallicity for the $\tau = 0.5\,$Gyr $e$-folding star formation
model. Although the spheroidal obscuration makes the ongoing star-formation
model slightly redder than a typical elliptical galaxy at low redshift, we
include it to approximate the effects of dust during the starburst
(i.e. PEGASE does not allow for the destruction of dust). We would expect this
model to be accurate at high redshift, but too red at low redshift.
Since the low redshift models are excluded in both cases, this is not a
concern. Shown for reference on each of the color redshift plots is the
line representing the color of a point source at the $5\,\sigma$ limiting
magnitude (AB) of the SDSS filter ($g_{\rm lim}=23.3$, $r_{\rm lim}=23.1$,
$i_{\rm lim}=22.3$) with the $K$-band magnitude assumed from a linear fit
to the $K-z$ diagram, given by: $K = 4.62\,\log(z)+17.2$, derived
from a fit to the galaxies in \citet[]{VanB:98}. Sources below the line
on the diagram would be detected in each SDSS band, while sources above
would not. Note that there will be some scatter in the cutoff due to the
scatter in the $K-z$ Hubble diagram. The exact nature of the objects passing
the magnitude cut will depend on the color dependence of the scatter
in th $K-z$ relation.  Combining the information in the $K-z$ diagram
and Figure~\ref{simplerk} we see that all of the optically bright
galaxies are either at low redshift, or galaxies very near their initial
formation redshift.  We can now use the optical properties of the radio
galaxies to eliminate the low redshift component of the distribution.

\begin{figure}
\begin{center}
\includegraphics[width=235pt]{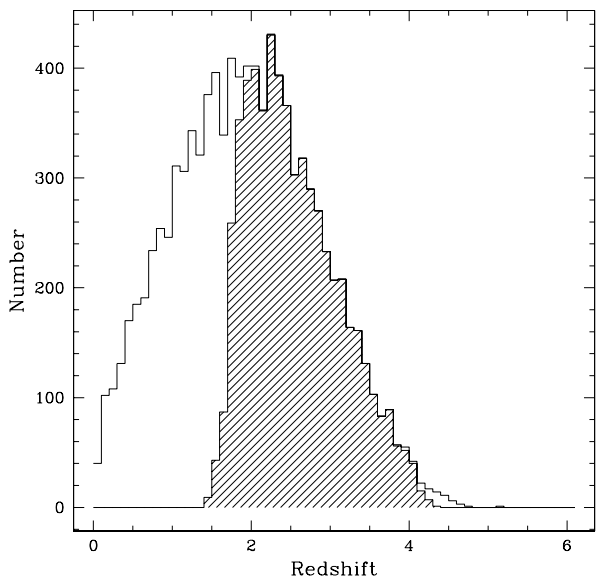}
\caption{Monte-Carlo realization of the $dN/dz$ distribution of Figure~\ref{dndz}.  The shaded histogram represents objects that pass our optical color cuts, assuming that all galaxies have a formation redshift $z_{\rm form}=5.0$.
\label{moncarlodndz}}
\end{center}
\end{figure}

Figure~\ref{moncarlodndz} shows a Monte-Carlo realization of the $dN/dz$
distribution in Figure~\ref{dndz}.  The shaded histogram represents the
galaxies that pass our selection criteria, assuming that all galaxies have
the colors of the solar metallicity instantaneous burst with $z_{\rm form}=5$.
These histograms show that using optical properties is an effective
way of screening out the low redshift component of the radio galaxy
population. The PEGASE models predict that the radio galaxies get much
bluer and brighter in the optical near their formation epoch due to
the initial burst of star formation in the models.  As the histogram
in Figure~\ref{moncarlodndz} shows, some fraction of these galaxies
will be excluded from optically selected samples near their formation
epoch as they become bright and blue enough to be detected in most optical
surveys. These model colors are especially sensitive to the assumptions made
for the star formation (e.g. obscuration, $e$-folding time), and it is
difficult to infer how accurate these color tracks will be near the initial
starburst. We also note that there is evidence for moderate obscuration in
some high redshift galaxies \citep[e.g.,][]{Dey:95,Ouch:04,Vil:05,Cha:05}.
Nevertheless, our models show that we may be somewhat biased against selecting
unobscured galaxies with ongoing star formation near their formation epoch.
Lowering the optical magnitude cutoff would allow these galaxies to enter our
sample, although at the expense of the low redshift cutoff.  Since we
are primarily interested in galaxies with evolved stellar populations
even at high redshift, as well as maintaining the efficiency of our search,
we retain the optical selection to favor the low redshift cutoff
at this minor expense of completeness.

\subsection{Selection using {\em SDSS} and {\em FIRST}}


We begin with the 2003 April 11 version of the FIRST catalog and select
all sources with integrated flux $S_{\rm 1.4GHz} > 100\,$mJy. 
The primary motivation for this cut is to define a manageable sample size.
Flux limits as low as $\sim$10 mJy are reasonable for selecting high-redshift
radio galaxies, and lower flux density limits will be explored in subsequent
work.  The $S_{\rm 1.4GHz} >100\,$mJy objects are positionally cross matched
with photometric data from SDSS Data Release 2 (DR2), which covers more than
3300 square degrees. Objects with candidate identifications in any of the
$u$, $g$, $r$, $i$ or $z$ bands within a conservative radius of $8''$ are
excluded. Radio sources with no cataloged SDSS counterpart are visually
inspected to further exclude possible low signal-to-noise optical counterparts,
as well as identifying and excluding extended or multiple-lobe radio sources
with an obvious optical counterpart located some distance from the cataloged
radio position (an optical source lying between independently cataloged radio
lobes, for example). Note that we are not restricted to unresolved radio
sources: Our selection criteria allow for extended and multicomponent radio
sources to be included as well.  To further eliminate likely low redshift
sources, we optimally combine \citep[]{Sza:99} the $g$, $r$, and $i$ SDSS
images (the three most sensitive of the five SDSS filters) and eliminate any
radio source with a candidate counterpart in the combined image. 

The $5\,\sigma$ point source limiting AB magnitudes for SDSS are
$u=22.3$, $g=23.3$, $r=23.1$, $i=22.3$, $z=20.8$, \citep[]{Ive:00}. The
coadded $g$, $r$, and $i$ SDSS images allow us to extend the low redshift
range being excluded by, in essence, improving our magnitude threshold by
$\approx 1.0\,$magnitude. This can be roughly thought of as providing an
effective $2\,\sigma$ $r$-band magnitude limit of $r\approx 24.1$, although
the specific limiting value is dependent on the details of
individual target galaxy SEDs. Coadded images are processed with
{\em SExtractor} version 2.3.2 \citep[]{Sex:96}, and any objects with a
detection greater than $2\,\sigma$ above the background are excluded from the
sample. Remaining target candidates are again visually inspected to eliminate
new (faint) optical identifications of extended or multi-lobe radio sources. 
This visual inspection introduces some subjectivity into the target selection 
criteria, but it is necessary in order to eliminate obvious SDSS counterparts 
to double lobed radio sources, and sources that are extended in the FIRST 
catalog.  Candidate targets with nearby bright stars or other nearby bright 
confusing sources are also excluded from the final target list. These targets 
are then checked against the NASA Extragalactic Database ({\em NED})\footnote{The NASA/IPAC Extragalactic Database (NED) is operated by the Jet Propulsion Laboratory, California Institute of Technology, under contract with the National Aeronautics and Space Administration.} to screen
out objects that had previously been observed. Only two objects,
4C $-$00.62 \citep{Rot:97} and 3C 257, \citep{Hew:91,VanB:98}
were previously identified with confirmed redshifts of $z=2.53$ and
$z=2.474$ respectively. The high redshift of these two radio galaxies
support the effectiveness of our selection criteria.  A final cut of
${\rm RA}> 12\,$hr is made to eliminate targets not visible during our
Spring observing runs.  This process yields a total of 172 target objects.
With 2085 $S_{\rm 1.4GHz} > 100\,$mJy radio sources in this area of DR2,
less than one in ten meet our selection criteria, giving us an order of
magnitude improvement in efficiency of finding high redshift radio galaxies
over blind spectroscopic targetting of all radio galaxies.

\begin{figure}
\begin{center}
\includegraphics[width=235pt]{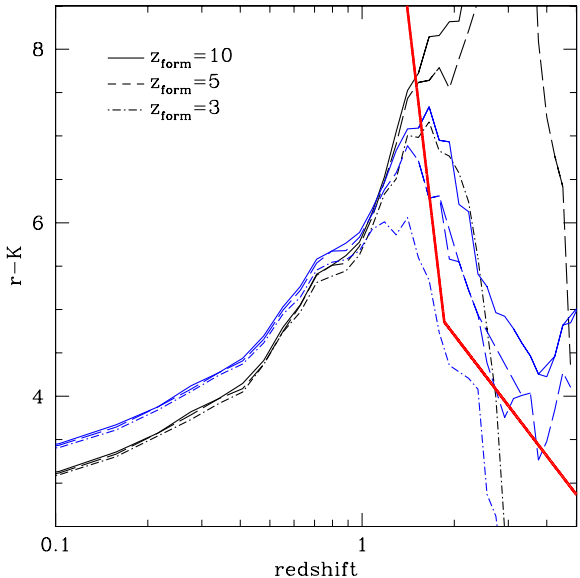}
\caption{Optical color cuts compared to model $r-K$ color as a function of redshift.  Objects above and to the right of the thick line pass the selection.
 \label{obsrk}}
\end{center}
\end{figure}

Figure~\ref{obsrk} shows a simulation of our multicolor optical selection
in terms of $r-K$ color as a function of redshift for the same two sets of
PEGASE spectral synthesis models used in Figure~\ref{simplerk}. The thick line
shows the approximate effect of our optical selection criteria based on
the $gri$ coaddition.  Using the model $dN/dz$ of Dunlop
and Peacock for radio sources brighter than $100\,$mJy with assumed pure
luminosity evolution, we expect to exclude all sources with $z\,<\,1.7$
and an average redshift for the sample of $z\,\sim\,2.5$.

\subsection{Selection on previous surveys}
As a test of our selection criteria, we apply our selection to subsets of
the galaxies in a blind radio survey \citep[]{Bes:99} and the ultra-steep
spectrum (USS) sample of \citet{Deb:01}.  Of the 178 radio sources with
spectroscopic redshifts in \citet[]{Bes:99}, only 20 fall within SDSS Data
Release 3 (DR3).  These sources have redshifts ranging from $z=0.004$
to $z=2.474$, a mean redshift of $z=0.78$, and a median of $z=0.66$.
Applying our selection criteria leaves only two targets at redshifts of
$z=1.339$ and $z=2.474$, the only galaxy with $z>1.5$ in the sample.
Only 9 of the 62 USS sources of \citet[]{Deb:01} are within SDSS DR3,
eight of which have firm redshifts.  The mean and median redshift of
this sample are 2.5 and 2.14 respectively.  Applying our selection
criteria eliminates the two $z<1$ galaxies from the sample, as well as
one unusually optically bright galaxy with $z=2.48$, where Ly$\alpha$ falls in the $g$ band, and \citet[]{Deb:01} note that it has
relatively strong continuum emission. 
The five remaining
galaxies have a mean and median of $z=2.58$.  As these two datasets show,
our selection criteria is very effective at eliminating the low redshift
foreground from both blind and USS samples.  The exclusion of one $z>2$ galaxy illustrates the fact that our selection 
may eliminate some galaxies with strong observed frame optical emission, and we will explore the extent of this bias in a future paper.  

\section{The Observations}
\label{obs}
We aquired $K$-band images for 96 of the 172 unique candidates on the KPNO
4-meter Mayall telescope over two sets of four nights:  2003 April 20-23
and 2004 May 31 to 2004 June 3, using the Florida Multi-object Imaging
Near-IR grism Observational Spectrometer ({\em FLAMINGOS}) instrument.
The detector is a 2048x2048 HgCdTe wide-field IR imager and multi-slit
spectrometer with a pixel size of $0.3165''$, which gives a $10.8'
\, \times \,  10.8'$ FOV on the 4 meter telescope.  Conditions were
highly variable during the 2003 run, with a night and a half lost due
a combination of high wind and moisture (first night seeing $1.0''$
to $1.5''$, subsequent nights $1.0''$ to $2.3''$ due to combination of
cloud, wind, and moisture).  The 2004 run was photometric on all four
nights with seeing varying from $0.7''$ to $1.2''$. Exposures were 20
seconds each in the 2003 run and 15 seconds each in 2004 to account
for brighter sky levels.  We observed in a fixed five point dither
pattern with a separation of 30 arcseconds.  Targets were observed for 15
minutes unless otherwise noted.  The $5\,\sigma$ limiting magnitude in
individual images spans $K\sim19.0$.  Three objects among those not
detected after fifteen minutes were observed a second time. Objects
{\em J1411+0124}, {\em J1350+0352}, and {\em J2242-0808} have total
integration times of 40, 61, and 29 minutes, respectively.
{\em J1123+0530} is the well known $z = 2.474$ radio galaxy
{\em 3C 257} \citep[]{Hew:91,VanB:98}, one of the most luminous radio
galaxies known.  As it passed our selection criteria, it was observed in
order to provide a consistency check with previous radio galaxy searches.

\subsection{Reduction and Astrometry}
\label{redux}
We processed the data using standard {\em NOAO IRAF}\footnote{IRAF is distributed by the National Optical Astronomy Observatories, which are operated by the Association of Universities for Research in Astronomy, Inc., under cooperative agreement with the National Science Foundation.} routines.
A set of dark frames were taken and subtracted from each image.
A set of 6-10 adjacent (in time) images were combined to create a
sky flat for subtraction (i.e. ``running sky flats").  The images
were approximately aligned based on the dither offsets, then the IRAF
tasks {\em mscgetcat} and {\em msccmatch} were used to accurately register
the images for coaddition.  The IRAF task {\em msccmatch} uses a catalog
of {\em USNO-A2} \citep[]{Mon:98} stellar positions and magnitudes for
image registration and transformation, which may include image shift,
scale change, and axis rotation.  The resulting astrometry displayed
a systematic offset on most images of between $0.5''$ to $1.0''$, due
to slight differences between the USNO-A2 and SDSS astrometry.  We manually
corrected for these systematic offsets when constructing catalogs of
each field.  Our final astrometry is accurate to subarcsecond precision,
with the residual difference between the $K$-band and SDSS positions well
fit by gaussian of width $0.25''$.

\subsection{Photometry}
\label{photom}
Each FLAMINGOS field was processed with {\em SExtractor} version 2.3.2.
Quoted $K$-band magnitudes are SExtractor MAG$\_$AUTO unless otherwise
noted.  Due to the large field of view of the FLAMINGOS instrument,
a large number ($\sim$100) of bright sources detected in the 2 Micron
All Sky Survey ({\em 2MASS})\footnote{This publication makes use of data products from the Two Micron All Sky Survey, which is a joint project of the University of Massachusetts and the Infrared Processing and Analysis Center/California Institute of Technology, funded by the National Aeronautics and Space Administration and the National Science Foundation.} are present in each field.  Sources were
cross matched with point sources from the 2MASS catalog having $K_{s} < 15.3$,
the $5 \, \sigma$ limiting magnitude for point sources. Saturated stars were
excluded from the comparison. A linear least squares fit between the
SExtractor and 2MASS objects was perfomed to determine the zero point offset
between the two datasets. This solution was assumed to continue linearly
beyond the $K_{s} = 15.3$ limit.  Although we observed in $K$-band, while
2MASS uses $K_s$, no significant color terms were evident (the magnitude
comparison between SExtractor and 2MASS was linear with a slope of unity).
For radio galaxies not detected in $K$-band at the $2\,\sigma$ level,
we report the $2\,\sigma$ magnitude of a point source as a lower limit.

\subsection{Star-Galaxy Separation and ERO definition}
\label{stargal}
The final $K$-band catalog was positionally cross matched with the SDSS
catalog using a webservice interface to OpenSkyQuery
(\verb+http://www.openskyquery.net+) in order to obtain $r-K$ colors.
A simple nearest neighbor criterion was used, and objects with no SDSS
counterpart within a three arcsecond radius were assigned a $5\,\sigma$
limiting magnitude of $r=23.1$.

Star-Galaxy separation was done in two steps:  For objects detected
in SDSS with $r\leq 21.0$ we used the SDSS star-galaxy classification
\citep[{\em probPSF}, which is described in][]{Scr:02}.  For fainter $r$-band
objects the separation was done by examining the difference between
the $K$-band MAG$\_$AUTO magnitude (from SExtractor) and a
fixed $3.5''$ aperture magnitude returned by SExtractor as a function
of MAG$\_$AUTO to separate pointlike objects from extended galaxies.

\begin{figure}
\begin{center}
\includegraphics[width=235pt]{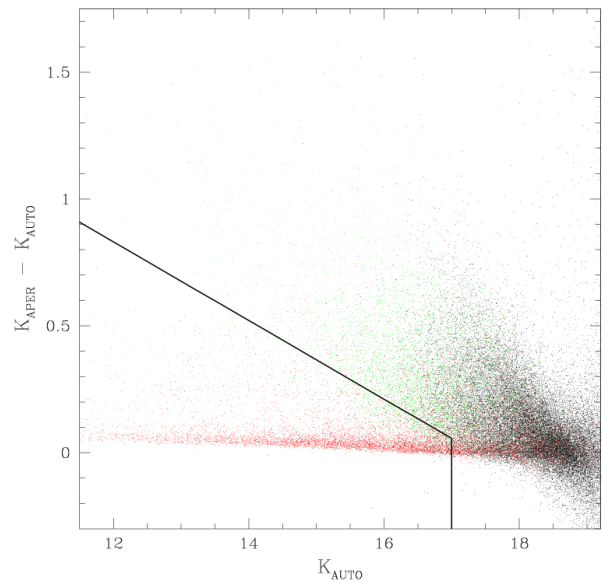}
\caption{ $K_{\rm APER(3.5'')}-K_{\rm AUTO}$  as a function of $K_{\rm AUTO}$.
Red and green points are $r<21.0$ objects classified by SDSS as stars and galaxies respectively.  Black points above and to the right of the black line are considered galaxies. 
 \label{sgfig}}
\end{center}
\end{figure}

Figure~\ref{sgfig} shows the resulting star-galaxy separation.  This
separation was chosen to be in rough agreement with the star-galaxy
classification from SDSS where reliable classification was available.
The slope of the resultant number counts for both stars and galaxies are
in good agreement with those of \citet[]{Dad:00}.  Reliable separation
was possible to $K=17.0$, and objects fainter than this are considered to
be galaxies.

The exact definition of EROs in the literature is not standardized.
Some authors use total magnitudes \citep[]{Wold:03,Cima:02}, some
isophotal/aperture corrected \citep[]{Dad:00}, some use matched aperture
magnitudes in the $R$- and $K$-bands\citep[e.g.][]{FLAM:06}, among others.
There is the additional variation in the use of specific $R$ and $K$
(and $K_s$) filters used to select EROs, making comparison between ERO
samples problematic.  We chose our ERO definition by comparing to publicly
available data in the Bo\"{o}tes field of the FLAMINGOS Extragalactic
({\em FLAMEX}) Survey \citep[]{FLAM:06}, which lies in the NOAO Deep
Wide Field Survey ({\em NDWFS})\footnote{The NOAO Deep Wide-Field Survey (Jannuzi and Dey 1999) is supported by the National Optical Astronomy Observatory (NOAO). NOAO is operated by AURA, Inc., under a cooperative agreement with the National Science Foundation.} \citep[]{Jan:99}, as well as SDSS.
The Bo\"{o}tes field of the FLAMEX survey covers approximately 4.7
square degrees in both $J$- and $K_s$-bands.  The FLAMEX survey was
conducted with the FLAMINGOS instrument, and the its location within
SDSS allows for a direct comparison with our data, although they use the
$K_s$ filter where we use $K$.  The FLAMEX survey defines EROs based on $6''$
matched aperture magnitudes in NDWFS $R$ and FLAMEX $K_s$ bands.
Unfortunately, the SDSS catalog does not include aperture magnitudes,
so we use SDSS model magnitudes in conjunction with $4''$ aperture
magnitudes in $K$-band to define our $r-K$ color, as this aperture appears
to best mimic the FLAMEX results.  Cross-matching the FLAMEX, SDSS, and NDWFS data in
a subset of the Bo\"{o}tes field, we found that an ERO definition of
$r-K_{\rm APER(4'')}\,\geq\,5.50$ approximated the
$R-K_{S}\,\geq\,5.0$ definition of \citet[]{FLAM:06}.

\section{Results}
\label{res}
\subsection{The Radio Galaxies}
The results of our $K$-band observations are shown in Tables\,\ref{detections} and \ref{nondetections}.  Table\,\ref{detections} lists objects detected, while Table\,\ref{nondetections} lists non-detections in the infrared.  The column entries are:
\begin{enumerate}
\item Object Name: Name of source in IAU J2000 format.  Objects with double lobe and multiple radio morphologies have the infrared object reported first and the individual radio components named a, b, c as necessary.
\item Date: Date observed.
\item Radio RA and DEC:  Position of radio detection from the FIRST catalog (J2000)
\item $K$-band RA and DEC:  Position of the infrared detected object (J2000)
\item Extended: Y if object shows visible extension in the FIRST image or if the object is a double or multiple lobe radio source.  N if object is unresolved in the radio image. 
\item $f_{\rm peak}$: Peak $1.4\,$GHz flux density from FIRST catalog (mJy)
\item $f_{\rm int}$:  Interated $1.4\,$GHz flux density from FIRST catalog (mJy)
\item $K$:  apparent magnitude and error for infrared object. Magnitudes are in the 2MASS system, which is slightly offset from the CIT system, given by: $K_s(\rm 2MASS) = K(\rm CIT) + 0.0000\,\times\,(J-K)_{\rm CIT} - 0.024 \pm 0.003$ \citep[]{Car:01}.  For nondetections (Table \ref{nondetections}) the $2\,\sigma$ detection limit is reported.  Errors are those output by SExtractor.
\item Seeing: The full width half maximum seeing measured for the field, given in arcseconds.
\item $z_{K-z}$: estimated redshift based on a fit to the Hubble $K-z$ relation, given by $K=4.62 \log(z) + 17.2$.  For nondetections (Table  \ref{nondetections}), the estimated redshift is given as a lower limit for the $2\,\sigma$ limiting magnitude.
\end{enumerate}

We observe a total of 96 unique objects, detecting 70 of these at a
$2\,\sigma$ level to a magnitude limit of $K\,\approx\,19.5$ in most cases.
Three objects, {\em J1102+0250}, {\em J1237+0135}, and {\em J1606+4751} show
highly extended double radio lobes with complex morphologies.  Subsequent 
visual inspection of the coadded $gri$ SDSS images for these objects reveals a
probable optical counterpart that was missed in the original target selection.
They thus do not meet our selection criteria, and we include them here
merely for completeness.  The total areal coverage of the $K$-band fields
(exluding the field of {\em J0742+3256}, where a transient problem with the detector
corrupted a portion of the field, but did not affect the radio galaxy target)
is 8570 square arcminutes (2.38 square degrees). Of the 96 target objects,
27 are double lobed or multiple radio sources, having multiple components in
the FIRST catalog. All but four of these multiple sources are identified in
the NIR. Ten additional radio galaxies appear to be resolved in the radio
postage images, and eight of these are identified with $K$-band counterparts.

\begin{figure*}
\epsscale{0.8}
\plotone{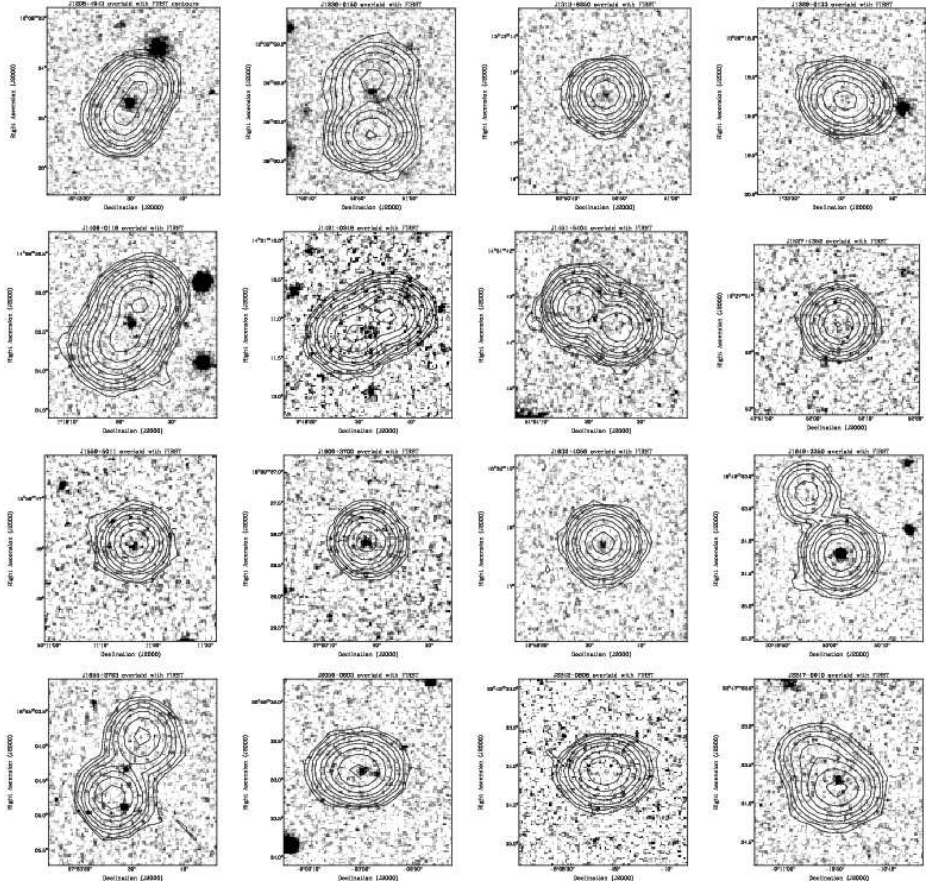}
\caption{A selection of $K$-band images.  North is to the right and East is down.  FIRST radio contours are overlaid with an outer contour at $5$mJy, and each subsequent contour indicating an increase by a factor of two.}
\label{prettypics}
\end{figure*}

Figure~\ref{prettypics} shows images for a selection of our $K$-band detected
objects. FIRST contours are overlaid with the outer contour at $5\,$mJy,
and each subsequent contour indicating an increase by a factor of two.
Postage stamp images for all galaxies can be found at
\verb+http://lahmu.phyast.pitt.edu/dragons/+.

\begin{figure}
\begin{center}
\includegraphics[width=235pt]{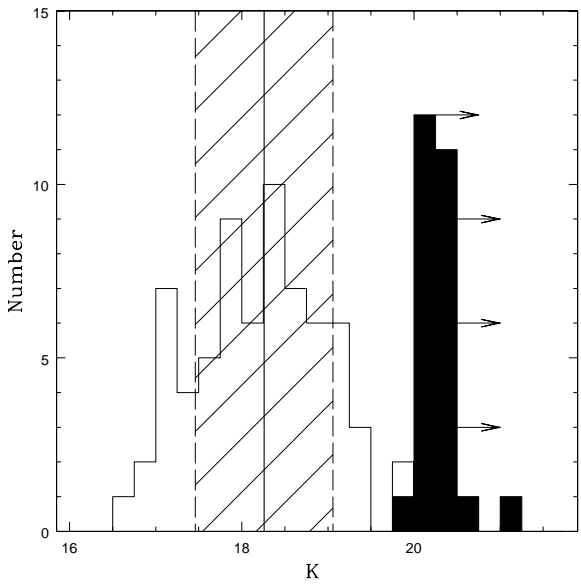}
\caption{Distribution of $K$-band magnitudes for the observed objects.
The filled histogram represents the $2\,\sigma$ limiting magnitudes of the
non-detections. The solid vertical line represents the $K$-band magnitude
for a galaxy at our expected lower redshift limit of $z=1.7$ and the shaded
area represents the approximate scatter given the $K-z$ diagram at
this redshift.
 \label{khist}}
\end{center}
\end{figure}
 
Figure~\ref{khist} shows the distribution of $K$-band magnitudes for
our observed objects.  The solid histogram represents $2\,\sigma$
lower limits for our non-detections.  The solid vertical line represents
the magnitude of an object at our expected redshift cutoff of $z=1.7$,
given our fit to the $K-z$ relation.  The shaded region represents
the approximate range of magnitudes for objects that follow the $K-z$
relation at this redshift.

\begin{figure}
\begin{center}
\includegraphics[width=235pt]{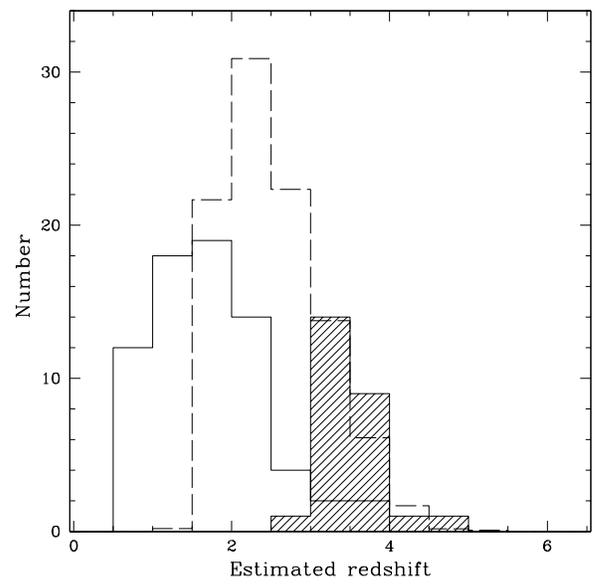}
\caption{Estimated redshift distribution based on a linear fit to
the $K-z$ Hubble diagram.  The shaded histogram represents lower limit
redshift estimates from the magnitude lower limits.  The dashed histogram
shows the distribution expected from convolving the \citet{DP:90}
$dN/dz$ with our selection criteria.
 \label{zhisto}}
\end{center}
\end{figure}

Figure~\ref{zhisto} shows the estimated redshift distribution based on
our linear fit to the $K-z$ Hubble diagram. Caution is advised in interpreting
this histogram, because of the large intrinsic scatter in the $K-z$ diagram.
The shaded portion of the histogram represents the lower limits on
the estimated redshifts for the non-detected objects.  The Monte-Carlo
simulation of the Dunlop and Peacock $dN/dz$ convolved with our selection
criteria from Figure~\ref{moncarlodndz} is shown for comparison.  If we
assign redshifts to the radio galaxies based on the linear fit to the $K-z$
diagram, the mean redshift for this sample is $z=2.5$ and the median
redshift is $z = 2.0$.  Comparison with the expected redshift distribution
in Figure~\ref{zhisto} shows that we have 35 objects with $K<18.3$ that we
expected to be excluded from our sample based on our model color tracks.
While three of these objects are resolved in the radio images and subsequent
closer examination revealed counterpart objects in SDSS, as indicated above,
a full third of our sample are brighter in $K$ than expected. This can be
partially attributed to the scatter in the $K-z$ relation if these galaxies
are at the upper end of the redshift range given their $K$ magnitude.
For example, the previously observed radio galaxy {\em 3C 257} (J1123+0530)
is known to be at $z=2.474$, while its $K=17.51$ magnitude corresponds to $z
\sim 1.2$ on a linear fit to the $K-z$ diagram, indicative of the large
intrinsic scatter in the $K-z$ relation.  \citet[]{VanB:98} point out
that $H \alpha$ falls into the $K$-band for {\em 3C 257}, which may
explain its relative brightness in $K$.  Similar line contamination may
be responsible for some of the brighter than expected objects in our sample,
but even this does not fully explain the brightest and reddest of our
objects. Thirteen objects with $K<17.5$ have anomalously red $g-K$,$r-K$,
and $i-K$ colors that are not fit by even our extreme model templates.
Galactic r-band extinction from the SDSS database for these objects are listed in Table~\ref{extinct}, based on the dust maps of \citet[]{Sch:98}.  There is a modest amount of extinction for three objects: J0941+0127, J1548+0036, and J1604-0013, though not enough to fully explain their extreme color.
Applying the extinction model of \citet[]{Cal:00} to the non-evolving
elliptical template of \citet[]{CWW:80} and assuming a redshift of near unity given
by the $K-z$ diagram, we require extinctions of $A_V > 0.5-1.5$ in order to
reach the lower limit $r-K$ colors of these infrared bright sources.
Determining spectroscopic redshifts for these objects will be the best
way to determine whether the anomalously red colors are due to obscuration
at lower redshift, a more luminous galaxy at higher redshift, or some
combination of the two. These objects could also be a separate class not
covered by our galaxy models, e.g. high redshift radio loud quasars or
low redshift type II quasars where the AGN is completely obscured.

\subsection{Radio Spectral Index}

Table\,\ref{specindex} shows the radio spectral index ($\alpha$, given by
$S_{\nu}\propto\nu^{-\alpha}$) computed from FIRST and the Texas $365\,$MHz
survey \citep[]{Dou:96}.  \citet[]{VanB:98} used an ultra-steep spectrum (USS)
cut of $\alpha^{1400}_{365} > 0.8$ to select a sample of high redshift
radio galaxy candidates. Of our 96 targets, 75 have observations at
$365\,$MHz, and of these 24 have $\alpha\,<\,0.8$ and would not be selected

\begin{figure}
\begin{center}
\includegraphics[width=235pt]{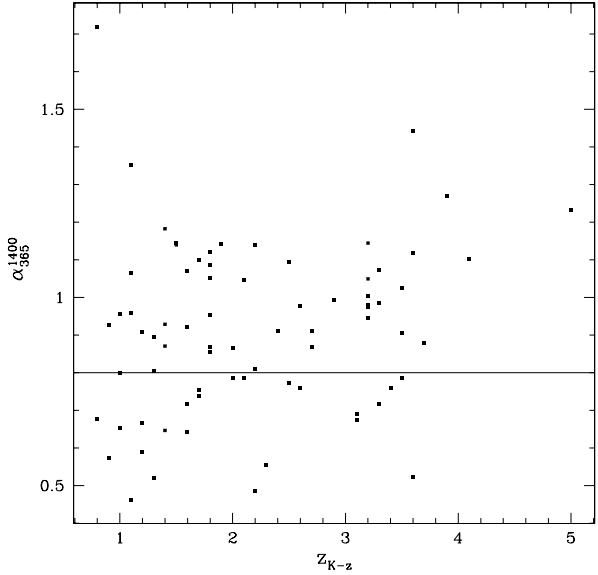}
\caption{Spectral slope as a function of $z_{K-z}$ for the 75 targets
with $S_{365\,MHz}$ flux densities measured in the Texas survey.
The horizontal line represents the USS cut of \citet[]{VanB:98}
 \label{alphred}}
\end{center}
\end{figure}

by the \citet[]{VanB:98} criteria.  Figure~\ref{alphred} shows the spectral

\begin{figure}
\begin{center}
\includegraphics[width=235pt]{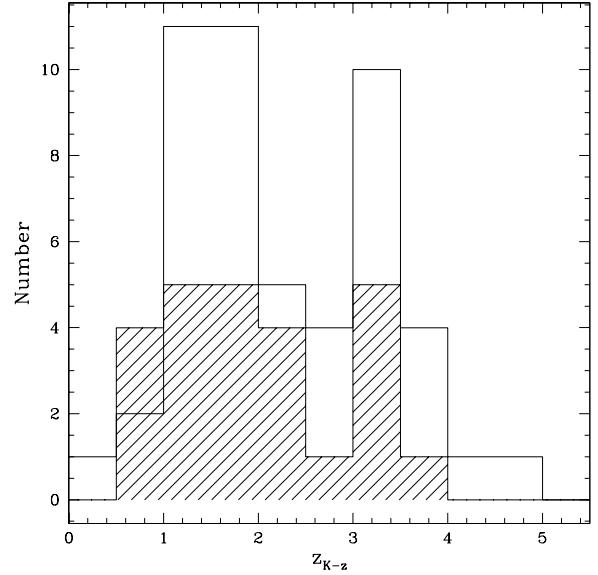}
\caption{Histogram of redshifts inferred from the $K-z$ diagram divided
into $\alpha\,<\,0.8$(shaded) and ultra-steep $\alpha\,>\,0.8$ emphasising
the presence at high redshift ($z_{K-z}>2$) of flatter spectrum sources.
 \label{usshist}}
\end{center}
\end{figure}

index of our targets as a function of $z_{K-z}$. Figure~\ref{usshist}
shows a histogram of the redshifts inferred from the $K-z$ diagram
for two spectral index ranges, flatter and steeper than $\alpha=0.8$.
The median redshift of the $\alpha\,<\,0.8$ sample is $z_{K-z}=2.0$ and
the mean is $z_{K-z}=2.13$, compared to a median of $z_{K-z}=2.0$ and
mean $z_{K-z}=2.17$ for the total sample, and median $z_{K-z}=2.0$ and mean
$z_{K-z}=2.30$ for the $\alpha\,>\,0.8$ sample. This suggests that flatter
spectral slope systems may contribute a non-negligible fraction of the HzRG
population. The insensitivity of our method to spectral slope allows us to
select candidates that would be missed by USS selection techniques. Applying
an USS criteria to our dataset would eliminate a full third of the targets.
\begin{figure}
\begin{center}
\includegraphics[width=235pt]{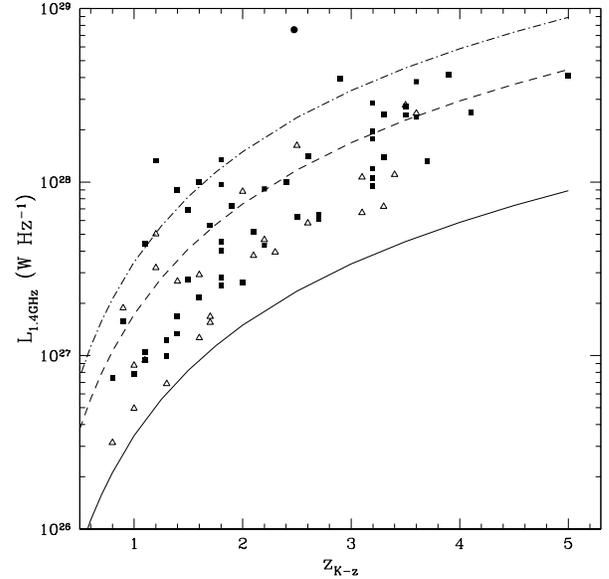}
\caption{$1.4\,$GHz radio luminosity as a function of $K-z$ redshift.
The lines represent $100\,$mJy (solid), $500\,$mJy (dashed), and $1\,$Jy (dot-dashed) flux density
limits assuming a spectral slope $\alpha = 0.4$. Squares mark objects
with $\alpha^{1400}_{365}>0.8$, triangles indicate
$\alpha^{1400}_{365}\leq 0.8$. The solid black circle marks the radio
luminosity of 3C~257 with spectroscopic redshift $z=2.474$.
 \label{radlum}}
\end{center}
\end{figure}
Figure~\ref{radlum} shows the $1.4\,GHz$ luminosities of the 75 galaxies
given their $\alpha^{1400}_{365}$ and $K-z$ redshift.  The three lines
represent $100\,mJy$, $500\,mJy$, and $1\,Jy$ $1.4\,GHz$ flux limits,
assuming $\alpha = 0.4$.

Table\,\ref{bigspecindex} lists a subset of our candidates with additional radio
observations at $4.85\,GHz$ \citep[From][]{Gre:91} and $151\,MHz$
\citep[From the 6C survey,][]{Hal:88,Hal:90}.  This table shows the
frequency dependence of the two point spectral slope.  Several of the
sources show significant deviations from a power law over the frequency
range in question, showing that the USS sample selection will differ
depending on frequencies used, while our selection method is unaffected
by objects with concave radio spectra.  

\subsection{Environment}

If, as we assert, high redshift radio galaxies form in the most
overdense regions of the early Universe, then hierarchical formation
scenarios predict an enhanced number of (proto-)galaxies associated
with this overdensity.  Although the strong spatial clustering of
EROs is often attributed to association with high redshift galaxy
overdensities, it is only recently that direct evidence for this has
been found \citep[]{Geo:05}.  We search for overdensities of both EROs
and $K$-band selected galaxies in the vicinity of our radio galaxy candidates.
We begin with a consideration of the source counts of both $K$-band objects, as well as EROs.
\begin{figure}
\begin{center}
\includegraphics[width=235pt]{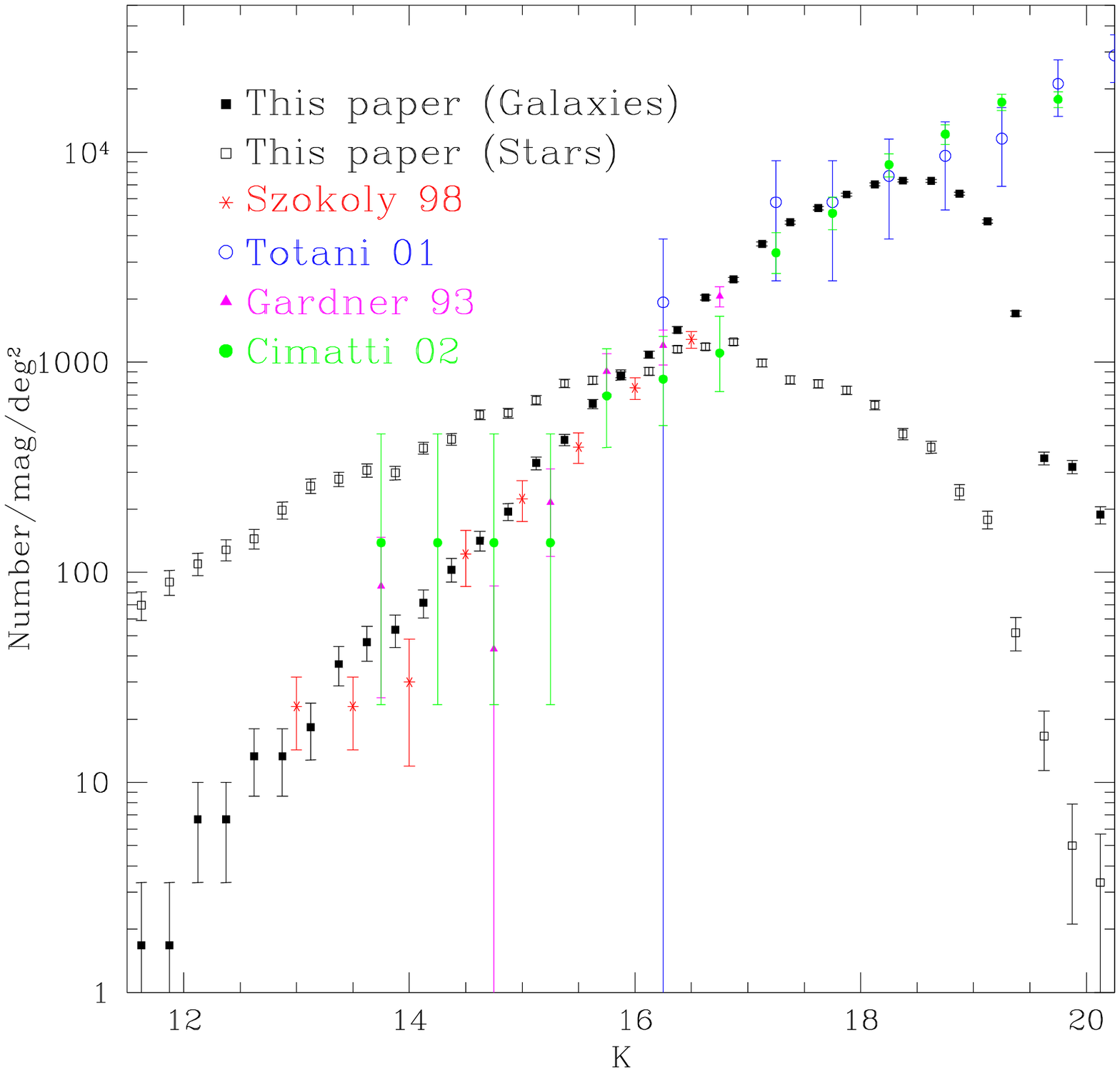}
\caption{Differential $K$-band galaxy number counts.  For comparison the number counts of \citet[]{Gar:93,Szo:98,Tot:01} and \citet[]{Cim:02} are shown.
 \label{numcnt}}
\end{center}
\end{figure}
Figure~\ref{numcnt} shows the $K$-band differential number counts for the
galaxies and stars in our 95 fields.  No completeness corrections have been applied.
Plotted for comparison are counts from \citet[]{Gar:93}, \citet[]{Szo:98}
($K_s$), \citet[]{Tot:01}, and \citet[]{Cima:02} (from the $K20$ survey,
also $K_s$).  The raw counts are given in Table\,\ref{numcounts}.  As the Figure
shows, we are in good agreement with previous infrared observations,
and it appears that the radio galaxy target selection does not bias our
$K$-band counts to the magnitude limits probed.
\begin{figure}
\begin{center}
\includegraphics[width=235pt]{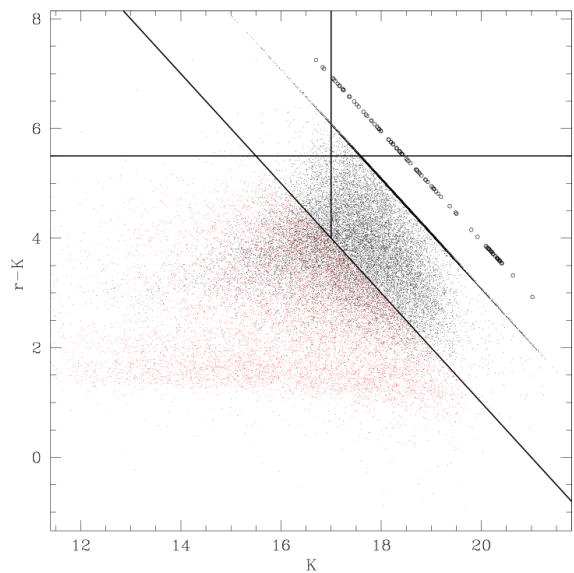}
\caption{$r-K$ as a function of $K$, showing star-galaxy separation
and ERO definition. The diagonal upper limit in the measurements is due to the $5\,\sigma$
$r$-band limit from the SDSS assigned to non-detections, while
the horizontal line marks the $r-K=5.50$ color used to define EROs.
Red points are objects classified as stars, while black are galaxies.
The diagonal line represents $r=21.0$, below which the SDSS
star-galaxy separation is used.  No star-galaxy separation is done for
objects above this diagonal line and to the right of the vertical line
at $K=17.0$. Open circles represent $2 \, \sigma$ $r-K$ lower limits for the target
radio galaxies in these fields.
 \label{newrkk}}
\end{center}
\end{figure}

Figure~\ref{newrkk} shows the $r-K$ as a function of $K$ color-magnitude
diagram after cross-matching our infrared catalog with the SDSS. We define an
Extremely Red Object (ERO) as having $r-K \geq 5.50$ as indicated by the
horizontal line on this Figure. Objects having no corresponding SDSS object
within $3''$ are assigned a limiting magnitude of $r=23.1$. This yields

\begin{figure}
\begin{center}
\includegraphics[width=235pt]{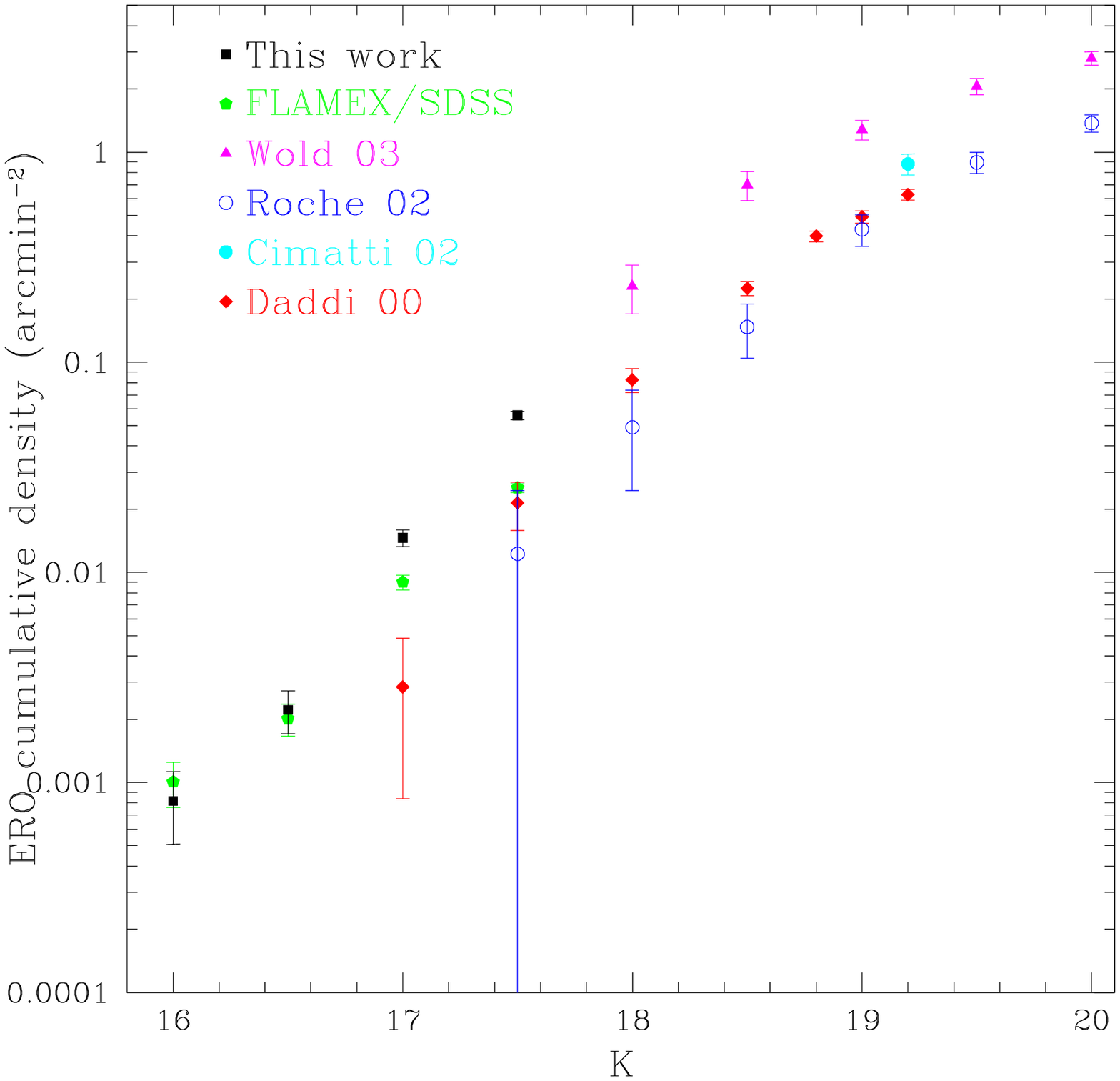}
\caption{Cumulative ERO surface density for our dataset and FLAMEX/SDSS.  Shown for comparison are data from \citet[]{Dad:00,Cima:02,Roc:02} and \citet[]{Wold:03}.
 \label{neweronumcnt}}
\end{center}
\end{figure}
a total of 479 EROs, and Figure~\ref{neweronumcnt} shows the cumulative
ERO surface density for this sample. Shown for comparison are the
SDSS/FLAMEX ERO cumulative density for the $\approx$ 4.7 degrees$^2$
of the Bo\"{o}tes field.  As can be seen, our ERO surface density
for $K\,>\,17.0$ EROs is higher than the surface densities of both
\citet[]{Dad:00} and the FLAMEX/SDSS data, suggestive of an overdensity
of EROs associated with our radio galaxy targets.  \citet[]{Wold:03}
observe EROs surrounding $z \sim 2$ radio loud quasars, and the Figure
shows a higher ERO surface density than the ``field" surveys of
\citet[]{Roc:02}, \citet[]{Dad:00}, and \citet[]{Cima:02}.  Our data
seem to show a continuation of the trend seen by  \citet[]{Wold:03}.
Because of the $r=23.1$ limit from SDSS we are limited to studying
$K \leq 17.60$ EROs.  Due to the small areas covered by previous ERO
studies, surface densities at this bright K magnitude are uncertain.
$5.9\%$ (30 out of 509) of our EROs are classified as stars, smaller
than the $\approx 9\%$ of \citet{Man:02} and \citet{Wold:03}, which may be
a concern.  At these bright $K$-band magnitudes, robust star galaxy
separation is increasingly important. We note, again, that our overall star
and galaxy counts are in good agreement with \citet[]{Dad:00} in the relevant
magnitude range, leading us to believe that our star-galaxy separation
is robust.  Figure~\ref{numcnt} shows that our $K$-band number counts
are consistent with previous surveys, yet our ERO density is higher
at $K>17.0$ than in the same surveys.  This is because only a small
percentage of the $K$-band selected galaxies are EROs, and the increase in
this rare population is small compared to the overall galaxy counts.
Thus, the EROs appear to trace an overdensity not readily apparent in
the $K$-band data alone, as pointed out by \citet{Wold:03}.

We now examine the clustering of EROs around our radio galaxies.
We restrict our analysis to $K\,\leq\,17.50$ EROs in order to remain
complete given the $r_{\rm lim} = 23.1$ magnitude limit of our SDSS data.
Because we expect the radio galaxy to be, by far, the most luminous galaxy
in its local environment, we expect to see a possible overdensity around
only the brightest of our targets. An overdensity of EROs with $K$-band
magnitudes brighter than the radio galaxy would be a possible indication
of a foreground structure.  Field by field, the density of EROs is quite
inhomogeneous, as expected.  More than half of the fields have no bright EROs
within $100''$ of the radio galaxy target, while seven fields have three or
more EROs within this radius.  The seven densest fields have radio galaxy
targets with magnitudes evenly spaced in the range $K=17.4-20.6$, showing
no trend of high ERO density with radio galaxy NIR brightness, contrary to
\begin{figure}
\begin{center}
\includegraphics[width=235pt]{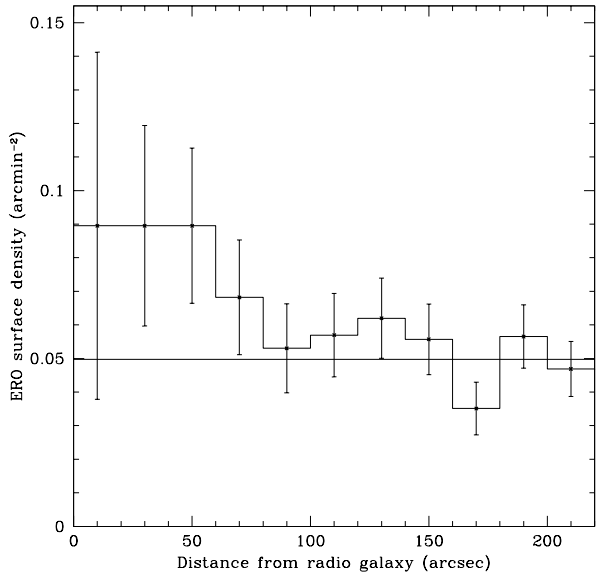}
\caption{Average surface density of $K<17.50$ EROs as a function of radial
distance from the radio galaxy for 95 target fields.  The four panels are
subsamples selected by the $K$-band magnitude of the target radio galaxy.
The horizontal line represents the ``local field density" of the sample,
defined as the average surface density of EROs between $120''$ and $180''$
from the radio source.
 \label{newerodist}}
\end{center}
\end{figure}
our expectations.  Figure~\ref{newerodist} shows the average distribution
of EROs with $K \leq 17.50$ as a function of radial distance from the
radio galaxy for 95 of our target fields.  We do not include the radio
galaxy itself, although some are EROs.  The horizontal line is a measure
of the ``local field'' density of EROs, defined as the density of EROs
between $120''$ and $180''$ from the radio source.  There is a modest
overdensity of EROs within an arcminute of the radio targets, although
the uncertainties are very large.  \citet[]{Wold:03} perform a similar
analysis and see no such evidence for clustering for EROs brighter
than $K_s\,=\,19.5$ around their $z\,\sim\,2$ radio loud quasars.
A deeper sample of EROs is necessary for a complete analysis of the
radio galaxy environment.  This could be established from our existing
data through the addition of deep $R$-band imaging ($R\,\approx\,25$),
and we are actively pursuing this goal.  Deep $R$-band imaging would
also allow for more rigorous star-galaxy separation to fainter magnitudes,
to better determine the stellar contamination of our ERO population.
\begin{figure}
\begin{center}
\includegraphics[width=235pt]{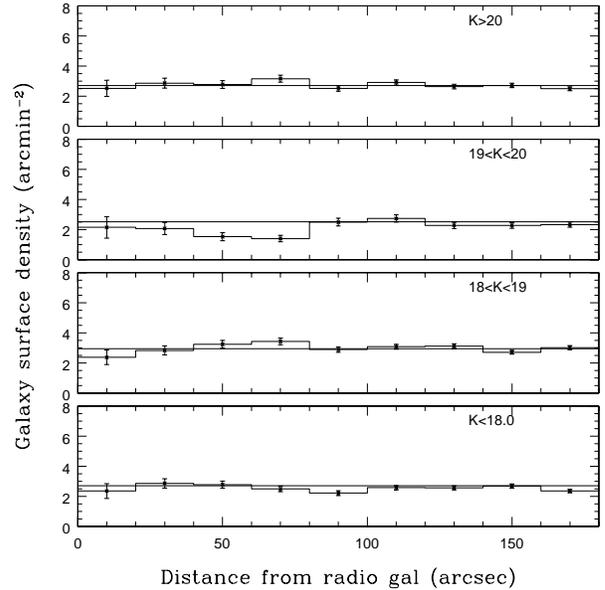}
\caption{Surface density of $K<19.0$ galaxies as a function of radial
distance from the radio galaxy.  The four panels are subsamples selected
by the $K$-band magnitude of the target radio galaxy.  The horizontal
line in each panel represents the ``local field density" of $K$-band selected
galaxies, defined as the average surface density of galaxies between $120''$
and $180''$ from the radio source.
 \label{kdist}}
\end{center}
\end{figure}

Figure~\ref{kdist} shows the distribution of $K\,<\,19.0$ galaxies around
our targets.  The four panels of Figure~\ref{kdist} represent the division
of the sample by the apparent magnitude of the target radio galaxy.
Given the $K-z$ Hubble diagram, this is a rough proxy for redshift.
The horizontal line in each panel shows a measure of the ''local field''
density of galaxies (defined as the density of objects between $120''$ and
$180''$ from the radio source).  No clustering is seen in any of the bins.
This is not unexpected, as \citet[]{Hal:98} and \citet[]{Hal:01} do not
see an excess around $z \approx 1.5$ radio loud quasars below $K=19$,
but do see an excess of galaxies at $K=19-20.5$.  Deeper $K$-band imaging
is necessary to detect possible companion galaxies at the high redshifts
of our radio galaxy targets.  Such imaging will be conducted on future
additions to our target sample (see \S\,\ref{summ} for details).
  
\section{Summary and Future Work}
\label{summ}
Using a novel radio-optical selection technique, we have obtained
NIR imaging for 96 HzRG candidates. Based on redshift estimates from the
$K-z$ Hubble diagram, our new selection technique appears to be effective
at identifying HzRGs, as more than half of our targets have $K$-band
photometry consistent with $z>2$. This technique is not sensitive to
radio spectral slope, and avoids the frequency dependence of USS techniques
for galaxies with non-power law radio spectra.  Of the 75 target galaxies
present in the Texas 365 MHz survey 24 have
$\alpha^{\rm 1.4GHz}_{\rm 365MHz} <0.8$. These would be excluded by the USS
criteria of \citet[]{VanB:98}, and all but four of these 75 have
$\alpha^{\rm 1.4GHz}_{\rm 365MHz} <1.3$, which would be excluded from
more recent USS selections \citep[e.g.][]{Deb:01,Jar:01}. Comparison with
the previous surveys of \citet[]{Bes:99} and \citet[]{Deb:01}
shows that our technique selects nearly all of their high redshift targets,
while also eliminating the low redshift sources. The $K$-band
number counts are consistent with previous work, and the large
non-contiguous area (2.38 square degrees) covered by the {\em DRaGONS}
survey to date makes it an excellent resource for exploring galaxy properties
through the combination of NIR and optical data.

We have uncovered a previously unseen class of radio
sources with anomalously red colors ($r-K\,>\,6.5-7$), which may be evidence of
significant obscuration at moderate redshifts.  These galaxies represent more than 
ten percent of our observed sample, indicating that they may be a substantial percentage of the radio galaxy
population.  Being non-detections in SDSS (with $r\,>\,24.1$), current optical AGN selection techniques are insensitive 
to these sources.  Radio loud QSOs comprise only $\sim\,5-10\%$ of overall QSOs; therefore, these 
objects could represent a significant contribution to radio loud AGNs
that are not counted in current samples.

We have identified a
sample of 479 EROs to a depth of $K=17.50$, one of the largest bright ERO
samples to date, and have identified a modest overdensity of EROs around
our HzRG candidates. We do not detect any similar overdensity when all the
$K$-band selected galaxies to $K=19.0$ are considered. 

Spectroscopic follow up of the radio galaxy candidates and the surrounding
EROs is of the utmost importance, both for validation of our HzRG candidates,
as well as determination of the relationship between the radio galaxies and
the ERO overdensity. Firm redshifts will also allow us to accurately test the $z-\alpha$ 
relation for our non-USS selected sample.  Near infrared spectroscopy of the anomalously
red radio galaxies should pinpoint the cause of their extreme colors, as well as 
determine the amount of obscuration present at intermediate and high redshifts.

Deeper optical imaging is required in order to
establish colors for the radio galaxies and explore their morphology in
more detail. Deeper optical imaging will also allow us to select a
deeper and more complete sample of EROs, allowing for a more detailed study
of the radio galaxy environment and the nature of the ERO overdensity. A
more quantitative study of ERO clustering will also be possible with deeper
optical data (including a two-point angular correlation function analysis).

A future paper will also examine the effects of both our radio flux density cut, as well as our optical color cuts.  In future observing runs we plan to target a small sample of lower flux density radio sources that meet our optical color cuts.  We will supplement this sample with galaxies from the FLAMEX survey that meet our selection criteria.  This dataset will allow us to examine any biases introduced by our radio selection.  Several of the targets in our most recent round of observing lie in the SDSS Southern Survey, an area of SDSS that is more than four times more sensitive than the standard SDSS survey.  This deeper imaging data will allow us to study the optical properties of those targets that are detected.



\acknowledgements

The authors would like to thank Julia Bryant for helpful input and discussion.

FLAMINGOS was designed and constructed by the IR instrumentation group (PI: R. Elston) at the University of Florida, Department of Astronomy, with support from NSF grant AST97-31180 and Kitt Peak National Observatory.

The SDSS is managed by the Astrophysical Research Consortium for the Participating Institutions. The Participating Institutions are the American Museum of Natural History, Astrophysical Institute Potsdam, University of Basel, Cambridge University, Case Western Reserve University, University of Chicago, Drexel University, Fermilab, the Institute for Advanced Study, the Japan Participation Group, Johns Hopkins University, the Joint Institute for Nuclear Astrophysics, the Kavli Institute for Particle Astrophysics and Cosmology, the Korean Scientist Group, the Chinese Academy of Sciences (LAMOST), Los Alamos National Laboratory, the Max-Planck-Institute for Astronomy (MPIA), the Max-Planck-Institute for Astrophysics (MPA), New Mexico State University, Ohio State University, University of Pittsburgh, University of Portsmouth, Princeton University, the United States Naval Observatory, and the University of Washington.

This research has made use of data obtained from or software provided by the US National Virtual Observatory, which is sponsored by the National Science Foundation.

SJS, AJC and AMH acknowledge partial
support from NSF ITR grant ACI0121671, NSF CAREER grant AST9984924, and NASA grant AISR NAG5-11996.

SJS would like to thank Belle \& Sebastian and Yo La Tengo for making music good enough to keep him awake at the telescope.

AMH acknowledges support from the Australian Research Council in the form of a Queen Elizabeth II Fellowship.


\clearpage
\LongTables 
\begin{landscape}
\begin{deluxetable}{rcllcccccccc}
\tablewidth{0pt}
\tablecolumns{12}
\tabletypesize{\tiny}
\tablecaption{Detections}
\tablehead{
\colhead{Object Name$\tablenotemark{a}$} & \colhead{date} & \colhead{radio RA} & \colhead{radio DEC} & \colhead{Extended?} & \colhead{K RA} & \colhead{K DEC} & \colhead{$f_{\rm peak}$(mJy)} & \colhead{$f_{\rm int}$(mJy)} & \colhead{K} & \colhead{seeing($''$)} & \colhead{$z_{K-z}\tablenotemark{b}$}
}
\startdata
J0742+3256 & 23 April 2003 & 07:42:13.61 & +32:56:51.16 & N & 07:42:13.60 & +32:56:51.20 & 130.51 & 144.60 & 17.72$\pm$0.08 & 1.0 & 1.3 \\ 
J0831+5210 & 23 April 2003 & \nodata & \nodata & Y & 08:31:44.83  & +52:10:34.95 & \nodata & \nodata & 18.93$\pm$0.17 & 1.1 & 2.4 \\ 
a  & \nodata &  08:31:44.89 & +52:10:41.28 & \nodata & \nodata &\nodata & 97.07 & 113.73 &   \nodata   &  \nodata & \nodata\\ 
b  &\nodata &  08:31:44.74 & +52:10:28.72 & \nodata & \nodata &\nodata & 108.06 & 131.18 &  \nodata    &  \nodata & \nodata \\ 
J0941+0127 & 23 April 2003 & 09:41:48.75 & +01:27:36.45 & N &09:41:48.77 & +01:27:35.89 & 117.82 & 120.56 & 17.17$\pm$0.08 & 2.2 & 1.0 \\ 
J0958+0324 & 20 April 2003 & \nodata & \nodata & Y &09:58:24.95 & +03:24:13.56 & \nodata & \nodata & 17.46$\pm$0.09 & 1.5 & 1.1 \\ 
a  &\nodata & 09:58:24.18 & +03:24:20.84 & \nodata &\nodata & \nodata & 72.53 & 190.15 &   \nodata  & \nodata & \nodata  \\ 
b  &\nodata &  09:58:23.57 & +03:24:24.98 & \nodata &\nodata & \nodata & 125.58 & 162.14 &  \nodata  & \nodata & \nodata  \\ 
c  & \nodata&  09:58:26.323 & +03:24:01.51 & \nodata & \nodata & \nodata & 166.91 & 286.56 &  \nodata  & \nodata & \nodata  \\ 
J1102+0250$\tablenotemark{c}$  &  21 April 2003 & \nodata & \nodata & Y & 11:02:06.59  & +02:50:45.45  &  \nodata &  \nodata & 16.70$\pm$0.06  & 2.1 &0.8 \\ 
a  &\nodata &  11:02:05.66 & +02:50:58.19 & \nodata & \nodata & \nodata & 29.48 & 127.69 & \nodata  & \nodata & \nodata  \\ 
b  &\nodata &  11:02:06.05 & +02:50:48.86 & \nodata & \nodata & \nodata & 8.71 & 33.29 & \nodata  &  \nodata & \nodata  \\ 
J1123+0530$\tablenotemark{d}$ & 23 April 2003 & \nodata & \nodata & Y & 11:23:09.43 & +05:30:18.60 & \nodata & \nodata & 17.51$\pm$0.11 & 1.7 & 1.2 \\ 
a  &\nodata &  11:23:09.06 & +05:30:20.58 & \nodata & \nodata & \nodata & 1514.62 & 1580.13 &  \nodata   & \nodata & \nodata \\ 
b  &\nodata  & 11:23:09.75 & +05:30:13.83 & \nodata & \nodata & \nodata & 150.31 & 163.07 &  \nodata   & \nodata & \nodata \\ 
J1135+0548 & 21 April 2003 &  11:35:17.81 & +05:48:54.13 & N & 11:35:17.84 & +05:48:54.06 & 190.84 & 212.63 & 18.77$\pm$0.15 & 2.1 &2.2 \\ 
J1155+0305 &  20 April 2003 & \nodata & \nodata & Y & 11:55:12.48 & +03:05:18.93 &  &  & 17.36$\pm$0.06 & 1.1 & 1.1 \\ 
a  &\nodata &  11:55:11.94 & +03:05:25.69& \nodata & \nodata & \nodata & 40.01 & 57.21 & \nodata  &\nodata  &\nodata  \\ 
b  &\nodata &  11:55:12.80 & +03:05:14.61& \nodata & \nodata & \nodata & 77.26 & 105.60 &  \nodata &  \nodata &\nodata \\ 
J1208+0414 & 21 April 2003 &  \nodata   &  \nodata  & Y & 12:08:11.90 & +04:14:59.37 & \nodata & \nodata & 18.15$\pm$0.14 & 1.7 & 1.6 \\ 
a  &\nodata &  12:08:11.69 & +04:15:04.87 & \nodata & \nodata & \nodata & 204.08 & 299.15 & \nodata    & \nodata & \nodata \\ 
b  &\nodata &  12:08:12.24 & +04:14:55.55 & \nodata & \nodata & \nodata & 264.65 & 341.17 &  \nodata   & \nodata & \nodata \\ 
J1208+4943 & 01 June 2004 & \nodata  &  \nodata  & Y & 12:08:24.73 & +49:43:29.72 & \nodata & \nodata & 17.03$\pm$0.04 & 0.8 & 0.9 \\ 
a  &\nodata &  12:08:24.04 & +49:43:31.74 & \nodata & \nodata & \nodata & 78.12 & 87.72 &  \nodata & \nodata  & \nodata \\ 
b  &\nodata &  12:08:24.94 & +49:43:28.61 & \nodata & \nodata & \nodata & 88.83 & 111.26 & \nodata   & \nodata & \nodata  \\ 
J1236+0150 & 02 June 2004 &  12:36:00.15 & +01:50:52.59 & Y &12:35:59.60 & +01:50:52.79 & 144.70 & 154.53 & 17.25$\pm$0.07 & 0.9 & 1.0 \\ 
J1237+0135$\tablenotemark{e}$ &  20 April 2003 & \nodata & \nodata & Y & 12:37:05.32 & +01:35:53.27 & \nodata & \nodata & 17.96$\pm$0.07 & 0.9 & 1.5 \\ 
a  &\nodata &  12:37:7.681 & +01:35:58.08 & \nodata & \nodata & \nodata & 162.2 & 193.99 & \nodata  & \nodata & \nodata  \\ 
b  &\nodata &  12:37:3.252 & +01:35:53.98 & \nodata & \nodata & \nodata & 180.7 & 232.25 & \nodata  &  \nodata & \nodata \\ 
J1250+6043 & 04 June 2004 &  12:50:24.47 & +60:43:46.84 & N & 12:50:24.47 & +60:43:46.59 & 288.67 & 304.05 & 17.81$\pm$0.09 & 1.1 &1.4 \\ 
J1259+0559 & 23 April 2003 &   \nodata  &  \nodata  & Y & 12:59:12.37 & +05:59:03.10 & \nodata & \nodata & 18.36$\pm$0.14 & 1.9 & 1.8 \\ 
a  &\nodata &  12:59:11.93 & +05:59:29.05 & \nodata & \nodata & \nodata & 66.15 & 73.60 & \nodata    & \nodata & \nodata \\ 
b  &\nodata  & 12:59:12.57 & +05:58:52.43 & \nodata & \nodata & \nodata & 53.98 & 105.19 &  \nodata   &\nodata & \nodata  \\ 
J1308$-$0022 & 03 June 2004 & \nodata & \nodata & Y &13:08:56.17 & $-$00:22:36.65 &  \nodata &  \nodata & 19.05$\pm$0.15 & 1.1 &2.5  \\ 
a  &\nodata &  13:08:56.36 & $-$00:22:33.12 & \nodata & \nodata & \nodata & 90.77 & 98.14 & \nodata  & \nodata & \nodata \\ 
b  &\nodata &  13:08:55.87 & $-$00:22:42.98 & \nodata & \nodata & \nodata & 134.01 & 143.03 & \nodata  & \nodata & \nodata \\ 
J1312+0009 & 20 April 2003 &  13:12:32.83 & +00:09:13.40 & N & 13:12:32.79 & +00:09:13.17 & 109.42 & 112.61 & 19.03$\pm$0.17 & 1.1 & 2.5 \\ 
J1313+6250 & 02 June 2004 &  13:13:15.63 & +62:50:47.29 & N & 13:13:15.63 & +62:50:47.36 & 128.32 & 132.18 & 18.43$\pm$0.10 & 1.1 & 1.8 \\ 
J1315+0533 & 04 June 2004 &  13:15:17.92 & +05:33:14.07 & N & 13:15:18.00 & +05:33:09.93 & 140.25 & 146.96 & 18.72$\pm$0.13 & 1.0 & 2.1 \\ 
J1332+0101 & 23 April 2003 & \nodata  & \nodata & Y & 13:32:16.55 & +01:01:48.26 & \nodata & \nodata & 17.05$\pm$0.08 & 2.0 & 0.9 \\
a   &\nodata &  13:32:16.78 & +01:01:50.83 & \nodata & \nodata & \nodata & 152.34 & 228.66 & \nodata  &  \nodata &  \nodata \\ 
b   &\nodata &  13:32:15.94 & +01:01:40.81 & \nodata & \nodata & \nodata & 118.60 & 184.03 & \nodata  &  \nodata &  \nodata \\ 
J1336+0207 & 03 June 2004 &  13:36:34.43 & +02:07:37.13 & Y & 13:36:34.33 & +02:07:42.21 & 113.55 & 127.76 & 19.20$\pm$0.18 & 0.9 & 2.7 \\ 
J1400+0053 & 21 April 2003 &  14:00:4.59 & +00:53:19.0 & N & 14:00:04.64 & +00:53:18.90 & 112.87 & 130.80 & 17.23$\pm$0.08 & 1.4 & 1.0 \\ 
J1402+0342 & 23 April 2003 &  14:02:24.86 & +03:42:27.10 & N & 14:02:24.89 & +03:42:26.73 & 501.26 & 540.42 & 17.65$\pm$0.10 & 1.7 & 1.2 \\ 
J1403+6048 & 04 June 2004 &  14:03:59.55 & +60:48:07.85 & N & 14:03:59.62 & +60:48:07.69 & 526.50 & 794.62 & 17.55$\pm$0.07 & 0.8 & 1.2 \\ 
J1408+0116 & 02 June 2004 &  \nodata   &  \nodata  & Y &14:08:33.38 & +01:16:22.24 & \nodata & \nodata & 17.08$\pm$0.05 & 0.8 & 0.9 \\ 
a  &\nodata &  14:08:33.13 & +01:16:23.99 & \nodata & \nodata & \nodata & 290.25 & 320.12 & \nodata  &  \nodata &  \nodata \\ 
b  &\nodata &  14:08:33.58 & +01:16:20.85 & \nodata & \nodata & \nodata & 199.15 & 292.63 &\nodata   & \nodata  &  \nodata \\ 
J1411+0124$\tablenotemark{f}$ & 20-21 April 2003 &  \nodata   &  \nodata  & Y & 14:11:08.29 & +01:24:40.56 & \nodata & \nodata & 19.93$\pm$0.18 & 1.3 & 3.9 \\ 
a  &\nodata &  14:11:08.32 & +01:24:45.48 & \nodata & \nodata & \nodata & 116.43 & 122.32 & \nodata  & \nodata & \nodata  \\ 
b  &\nodata &  14:11:08.21 & +01:24:33.78 & \nodata & \nodata & \nodata & 60.89 & 64.17 & \nodata  & \nodata & \nodata  \\ 
J1423+0139 & 23 April 2003 &  14:23:03.45 & +01:39:58.23 & N & 14:23:03.47 & +01:39:58.46 & 202.21 & 211.64 & 17.37$\pm$0.09 & 2.0 & 1.1 \\ 
J1438+0150 & 20 April 2003 &  14:38:17.15 & +01:50:31.05 & N & 14:38:17.20 & +01:50:31.27 & 83.24 & 118.52 & 17.94$\pm$0.17 & 1.0 & 1.4 \\ 
J1438+6149 & 03 June 2004 &  14:38:41.83 & +61:49:33.79 & N & 14:38:41.87 & +61:49:33.67 & 114.73 & 121.27 & 18.70$\pm$0.14 & 0.8 & 2.1 \\ 
J1451+5404 & 01 June 2004 &  \nodata   &  \nodata  & Y &14:51:43.43 & +54:04:22.02 & \nodata & \nodata & 19.37$\pm$0.20 & 0.9 & 2.9 \\ 
a  &\nodata &  14:51:43.66 & +54:04:25.67 & \nodata & \nodata & \nodata & 265.89 & 274.33 & \nodata  & \nodata & \nodata  \\ 
b  &\nodata &  14:51:43.23 & +54:04:25.67 & \nodata & \nodata & \nodata & 271.59 & 280.30 & \nodata  & \nodata & \nodata  \\ 
J1452+0032 & 21 April 2003 & \nodata & \nodata & Y & 14:52:00.78 & +00:32:45.54 & \nodata & \nodata & 17.92$\pm$0.09 & 1.3 & 1.4 \\ 
a  &\nodata &  14:52:01.46 & +00:33:01.29 & \nodata & \nodata & \nodata & 112.59 & 143.03 &\nodata     &\nodata &\nodata  \\ 
b  &\nodata &  14:51:59.91 & +00:32:41.68 & \nodata & \nodata & \nodata & 471.04 & 496.02 & \nodata    &\nodata &\nodata  \\ 
J1510+5244 & 03 June 2004 &  15:10:20.20 & +52:44:30.20 & N & 15:10:20.16 & +52:44:30.68 & 500.50 & 505.76 & 18.41$\pm$0.18 & 0.8 & 1.8 \\ 
J1515+5744 & 20 April 2003 &  15:15:29.30 & +57:44:57.32 & Y & 15:15:29.33 & +57:44:56.25 & 78.56 & 144.16 & 19.14$\pm$0.18 & 1.0 & 2.6 \\ 
J1523$-$0018 & 01 June 2004 &  15:23:16.19 & $-$00:18:55.40 & Y & 15:23:16.23 & $-$00:18:55.08 & 216.42 & 231.44 & 18.75$\pm$0.10 & 0.9 & 2.2 \\ 
J1526+0408 & 23 April 2003 & \nodata & \nodata & Y &15:26:37.15 & +04:08:14.22 & \nodata & \nodata & 17.87$\pm$0.11 & 2.0 & 1.4 \\ 
a  &\nodata  & 15:26:38.90 & +04:08:02.96 & \nodata & \nodata & \nodata & 18.14 & 53.66 &  \nodata   & \nodata & \nodata \\ 
b  &\nodata  & 15:26:36.59 & +04:08:19.72 & \nodata & \nodata & \nodata & 43.55 & 102.76 & \nodata    & \nodata & \nodata \\ 
J1532+4432 & 04 June 2004 &  \nodata   & \nodata   & Y &15:32:50.13 & +44:32:15.24 & \nodata & \nodata & 17.16$\pm$0.08 & 0.9 & 1.0 \\ 
a  &\nodata &  15:32:50.85 & +44:32:16.86 & \nodata & \nodata & \nodata & 92.53 & 117.98 & \nodata  & \nodata & \nodata  \\ 
b  &\nodata &  15:32:49.66 & +44:32:14.50 & \nodata & \nodata & \nodata & 86.96 & 131.67 & \nodata  & \nodata & \nodata  \\ 
J1541+5259 & 23 April 2003 & \nodata & \nodata & Y & 15:41:18.87 & +52:59:55.38 & \nodata & \nodata & 17.24$\pm$0.08 & 2.3 & 1.0 \\ 
a  &\nodata &  15:41:18.75 & +52:59:52.35 & \nodata & \nodata & \nodata & 100.27 & 115.12 &  \nodata   &\nodata &\nodata  \\ 
b  &\nodata &  15:41:18.81 & +53:00:00.38 & \nodata & \nodata & \nodata & 67.11 & 77.90 &  \nodata   &\nodata &\nodata  \\ 
J1543+5711 & 03 June 2004 &  15:43:30.33 & +57:11:32.40 & Y & 15:43:30.12 & +57:11:32.37 & 55.88 & 103.39 & 18.24$\pm$0.08 & 0.9 & 1.7 \\ 
J1547+4839 & 04 June 2004 &  15:47:42.06 & +48:39:12.20 & N & 15:47:42.34 & +48:39:09.33 & 186.86 & 214.75 & 19.49$\pm$0.18 & 0.9 & 3.1 \\ 
J1548+0036 & 21 April 2003 &  15:48:16.21 & +00:36:13.56 & Y & 15:48:16.22 & +00:36:13.27 & 123.08 & 126.05 & 16.83$\pm$ 0.04 & 1.7 & 0.8 \\ 
J1548$-$0033 & 20 April 2003 & \nodata & \nodata & Y & 15:48:21.75 & $-$00:34:00.80 & \nodata & \nodata & 19.06$\pm$0.14 & 0.9 & 2.5 \\ 
a  &\nodata &  15:48:21.45 & $-$00:33:59.28 & \nodata & \nodata & \nodata & 155.03 & 231.62 & \nodata  & \nodata & \nodata   \\ 
b  &\nodata  & 15:48:22.14 & $-$00:34:00.64 & \nodata & \nodata & \nodata & 157.66 & 201.46 & \nodata  & \nodata & \nodata  \\ 
J1549+4719 & 04 June 2004 &  15:49:53.47 & +47:19:48.56 & N & 15:49:53.53 & +47:19:48.25 & 104.66 & 106.33 & 18.20$\pm$0.11 & 0.9 & 1.6 \\ 
J1554+4729 & 04 June 2004 &  15:54:25.63 & +47:29:00.97 & N & 15:54:25.67 & +47:29:00.56 & 137.54 & 149.73 & 18.70$\pm$0.09 & 0.9 & 2.1 \\ 
J1559+5011 & 03 June 2004 &  15:59:47.89 & +50:11:16.41 & N & 15:59:47.96 & +50:11:16.44 & 127.0 & 130.28 & 18.38$\pm$0.11 & 0.8 & 1.8 \\ 
J1604$-$0013 & 02 June 2004 &  16:04:12.71 & $-$00:13:41.89 & N & 16:04:12.62 & $-$00:13:41.28 & 117.91 & 128.28 & 17.13$\pm$0.05 & 0.9 & 1.0 \\ 
J1604+4746 & 22 April 2003 &  16:04:27.85 & +47:46:35.02 & N & 16:04:27.92 & +47:46:34.48 & 351.96 & 366.10 & 18.70$\pm$0.13 & 1.3 & 2.1 \\ 
J1606+4751$\tablenotemark{g}$ & 04 June 2004 &  16:06:1.29 & +47:51:52.17 & Y & 16:06:01.40 & +47:51:46.30 & 3.07 & 12.31  & 17.47$\pm$0.07 & 0.9 & 1.1 \\ 
a  &\nodata &  16:06:1.82 & +47:52:01.68 & \nodata & \nodata &  \nodata & 36.38 & 61.54 &  \nodata  &  \nodata &  \nodata  \\ 
b  &\nodata &  16:06:2.49 & +47:51:21.20 & \nodata & \nodata &  \nodata & 81.73 & 109.02 &  \nodata  &  \nodata &  \nodata  \\
J1609+3700 & 03 June 2004 &  16:09:28.11 & +37:00:18.12 & N & 16:09:28.13 & +37:00:18.0 & 96.52 & 102.52 & 17.80$\pm$0.08 & 0.8 & 1.3 \\ 
J1617+4848 & 23 April 2003 &  16:17:25.42 & +48:48:28.69 & N & 16:17:25.47 & +48:48:28.38 & 236.95 & 244.32 & 18.00$\pm$0.11 & 1.9 & 1.5 \\ 
J1629+4937 & 20 April 2003 &  16:29:21.38 & +49:37:54.86 & N & 16:29:21.31 & +49:37:54.27 & 183.38 & 197.05 & 19.50$\pm$0.19 & 1.0 & 3.1 \\ 
J1632+4056 & 01 June 2004 &  16:32:16.27 & +40:56:32.71 & N & 16:32:16.29 & +40:56:32.63 & 200.82 & 203.15 & 18.89$\pm$0.13 & 0.8 & 2.3 \\ 
J1634+4155 & 03 June 2004 &  16:34:43.60 & +41:55:03.40 & N & 16:34:43.57 & +41:55:03.09 & 216.39 & 263.47 & 18.30$\pm$0.11 & 0.8 & 1.7 \\ 
J1636+4808 & 21 April 2003 & \nodata & \nodata & Y & 16:36:15.70 & +48:08:48.32 & \nodata & \nodata & 19.10$\pm$0.17 & 1.2 & 2.6 \\ 
a  &\nodata &  16:36:16.04 & +48:08:50.11 & \nodata & \nodata & \nodata & 87.46 & 97.73 &  \nodata   & \nodata & \nodata \\ 
b  &\nodata &  16:36:15.16 & +48:08:45.59 & \nodata & \nodata & \nodata & 147.42 & 165.66 &  \nodata   &\nodata  & \nodata \\ 
J1637+3223 & 03 June 2004 &  16:37:34.53 & +32:23:05.84 & N & 16:37:34.53 & +32:23:06.21 & 136.29 & 147.47 & 18.80$\pm$0.15 & 0.8 & 2.2 \\ 
J1643+4518 & 03 June 2004 &  16:43:22.62 & +45:18:06.68 & N &  16:43:22.79  & +45:18:01.68  & 108.30 & 110.39 & 18.30$\pm$0.11 & 0.8 & 1.7 \\ 
J1645+4152 & 03 June 2004 &  16:45:00.73 & +41:52:14.20 & N &16:45:00.80 & +41:52:14.30 & 112.56 & 115.96 & 18.42$\pm$0.09 & 0.8 & 1.8 \\ 
J1648+4233 & 23 April 2003 & 16:48:31.51 & +42:33:22.42 & Y & 16:48:31.52 & +42:33:21.82 & 167.44 & 169.30 & 19.00$\pm$0.17 & 2.1 & 2.5 \\ 
J1649+3350 & 02 June 2004 &  16:49:24.21 & +33:50:02.04 & N & 16:49:24.19 & +33:50:02.17 & 162.25 & 163.91 & 16.86$\pm$0.03 & 0.9 & 0.8 \\ 
J1654+4125 & 02 June 2004 &  16:54:43.87 & +41:25:02.96 & N & 16:54:43.89 & +41:25:03.00 & 225.60 & 228.56 & 18.15$\pm$0.10 & 0.9 & 1.6 \\ 
J1655+2723 & 01 June 2004 & \nodata & \nodata & Y & 16:55:04.33 & +27:23:28.91 & \nodata & \nodata & 18.52$\pm$0.09 & 0.8 & 1.9 \\ 
a  &\nodata &  16:55:03.87 & +27:23:32.00 & \nodata & \nodata & \nodata & 40.01 & 54.13 & \nodata  &\nodata &\nodata   \\ 
b  &\nodata &  16:55:04.73 & +23:23:26.47 & \nodata & \nodata & \nodata & 101.27 & 115.40 & \nodata  & \nodata &\nodata  \\ 
J1656+2707 & 02 June 2004 &  16:56:16.29 & +27:07:32.47 & N & 16:56:16.31 & +27:07:33.05 & 160.33 & 164.76 & 18.88$\pm$0.15 & 0.9 & 2.3 \\ 
J1707+2408 & 01 June 2004 & 17:07:44.58 & +24:08:54.56 & N & 17:07:44.57 & +24:08:54.41 & 144.83 & 169.15 & 17.96$\pm$0.09 & 0.8 & 1.5 \\ 
J1715+3027 & 01 June 2004 &  17:15:48.29 & +30:27:23.18 & N & 17:15:48.31 & +30:27:23.18 & 378.27 & 385.34 & 18.54$\pm$0.09 & 0.8 & 2.0 \\ 
J2059$-$0603 & 02 June 2004 &  20:59:32.88 & $-$06:03:00.34 & N & 20:59:32.87 & $-$06:02:59.59 & 155.63 & 161.72 & 18.32$\pm$0.10 & 0.9 & 1.7 \\ 
J2107$-$0701 & 02 June 2004 &  21:07:45.46 & $-$07:01:07.83 & N & 21:07:45.47 & $-$07:01:06.54 & 523.60 & 550.60 & 18.37$\pm$0.09 & 0.9 & 1.8 \\
J2223$-$0757 & 01 June 2004 &  22:23:26.52 & $-$07:57:08.07 & N & 22:23:26.51 & $-$07:57:06.94 & 104.79 & 108.34 & 17.70$\pm$0.08 & 0.9 & 1.3 \\ 
J2247$-$0910 & 01 June 2004 &  22:47:23.79 & $-$09:10:49.74 & Y & 22:47:23.69 & $-$09:10:49.26 & 77.10 & 103.37 & 17.99$\pm$0.09 & 0.9 & 1.5 \\ 
J2309$-$0834 & 03 June 2004 &  23:09:04.29 & $-$08:34:57.19 & N & 23:09:4.29 & $-$08:34:56.83 & 103.19 & 111.67 & 18.23$\pm$0.12 & 0.9 & 1.7 \\ 
J2316$-$0846 & 03 June 2004 &  23:16:35.08 & $-$08:46:17.72 & N & 23:16:35.07 & $-$08:46:17.16 & 101.85 & 105.40 & 18.59$\pm$0.14 & 0.9 & 2.0 \\ 
J2336$-$0838 & 04 June 2004 &  \nodata   &  \nodata  & Y & 23:36:18.35 & $-$08:38:48.87 & \nodata & \nodata & 18.50$\pm$0.15 & 0.9 & 1.9 \\ 
a  &\nodata &  23:36:18.11 & $-$08:38:43.58 & \nodata & \nodata & \nodata & 124.78 & 134.50 & \nodata  & \nodata & \nodata  \\ 
b  &\nodata &  23:36:18.65 & $-$08:38:48.27 & \nodata & \nodata & \nodata & 104.16 & 110.32 &\nodata   &\nodata  & \nodata  \\ 
J2337$-$0852 & 04 June 2004 &  23:37:32.44 & $-$08:52:39.49 & N & 23:37:32.46 & $-$08:52:38.83 & 118.04 & 120.14 & 18.20$\pm$0.13 & 0.9 & 1.6
\enddata
\tablenotetext{a}{Objects named with a and b are double lobed sources with the radio properties listed corresponding to
the object above them on the table}
\tablenotetext{b}{Based on a fit to the $K-z$ Hubble diagram $ K = 4.62* \log (z) + 17.2$}
\tablenotetext{c}{This object is an extended double lobe with a corresponding faint match in SDSS and should not have been included in our sample.  The object has SDSS magnitudes g=23.25 r=22.29 i=21.26 z=19.90}
\tablenotetext{d}{This object, {\em 3C 257} was previously identified as a $z= 2.474$ radio galaxy \citep[]{Hew:91}.  The $K$-band contains H$\alpha$ emission, leading to the underestimate of redshift from the K-z diagram \citep[see][for details]{VanB:98}.}
\tablenotetext{e}{This object is a wide separation double lobe with several candidate objects along the radio axis.  The probable match is faintly detected in SDSS with magnitudes of g=22.6 r=22.13 i=21.32 z=20.27}
\tablenotetext{f}{Object J1411+0124 was observed on subsequent nights for a total integration time of 40 minutes}
\tablenotetext{g}{This object has a match in SDSS g=22.86 r=22.45 i=22.11 z=21.12, was improperly targeted because it is a weak double lobe}
\label{detections}
\end{deluxetable}
\clearpage
\end{landscape}

\begin{deluxetable}{rcllcccccccc}
\tablewidth{0pt}
\tablecolumns{10}
\tabletypesize{\tiny}
\tablecaption{Non-Detections}
\tablehead{
\colhead{Object Name} & \colhead{date} & \colhead{radio RA} & \colhead{radio DEC} & \colhead{Extended?} & \colhead{$f_{\rm peak}$(mJy)} & \colhead{$f_{\rm int}$(mJy)} & \colhead{K ($2 \, \sigma$)} & \colhead{seeing($''$)} & \colhead{$z_{K-z}$}
}
\startdata
J1022+0357 & 23 April 2003 & 10:22:01.03 & +03:57:37.56 & N & 194.31 & 200.86 & $>$20.11  & 2.3 & $>$3.2 \\ 
J1028+0144 & 21 April 2003 & 10:28:02.79 & +01:44:06.51 & Y & 56.82 & 114.81 & $>$19.80 & 2.0 & $>$2.7 \\ 
J1044+0538 & 20 April 2003 & 10:44:19.88 & +05:38:07.98 & N & 125.54 & 130.20 & $>$20.32 & 1.2  & $>$3.5 \\ 
J1047+0216 & 23 April 2003 & 10:47:11.32 & +02:16:28.21 & N & 122.32 & 125.81 & $>$20.09  & 2.2 & $>$3.1 \\ 
J1144+0254 & 23 April 2003 & 11:44:34.26 & +02:54:25.56 & N & 114.01 & 119.18 & $>$20.13  & 2.1 & $>$3.2 \\ 
J1221+0248 & 23 April 2003 & 12:21:39.93 & +02:48:28.01 & N & 139.93 & 142.87 & $>$20.13  & 1.9 & $>$3.2 \\ 
J1234+0024 & 03 June 2004 &  12:34:30.79 & +00:24:59.45 & N & 94.84 & 100.55 & $>$20.35  & 1.2 & $>$3.6 \\ 
J1240$-$0017 & 21 April 2003 & 12:40:12.23 & $-$00:17:30.34 & N & 137.73 & 150.90 & $>$20.28  & 2.0 & $>$3.4 \\ 
J1303+0026 & 23 April 2003 & 13:03:57.48 & +00:26:45.41 & N & 91.23 & 104.04 & $>$20.15  & 2.0 & $>$3.2 \\ 
J1314+0330 & 21 April 2003 & 13:14:22.82 & +03:30:22.14 & Y & 163.59 & 250.05 & $>$20.31  & 1.6 & $>$3.5 \\ 
J1329+0133 & 01 June 2004 & 13:29:18.78 & +01:33:40.80 & N & 87.80 & 102.64 & $>$20.38 & 1.3 & $>$3.6  \\ 
J1350+0352$\tablenotemark{a}$ & 2003-2004 &  13:50:24.37 & +03:52:43.90 & N & 99.51 & 104.28 &$>$ 21.02& 1.1 & $>$5.0 \\ 
J1421+0248 & 01 June 2004 &  \nodata  & \nodata  & Y & \nodata   &\nodata & $>$20.33  & 1.0 & $>$3.5 \\ 
a  &\nodata & 14:21:10.957 & +02:48:35.76 & \nodata & 152.15 & 165.01 & \nodata  &  \nodata &  \nodata  \\ 
b  &\nodata & 14:21:11.20 & +02:48:29.10 & \nodata & 158.20 & 177.62 & \nodata  &  \nodata &  \nodata  \\ 
J1431+0511 & 04 June 2004 &  14:31:09.58 & +05:11:17.85 & Y & 1.31 & 2.22 & $>$20.19 & 0.9 & $>$3.3 \\ 
a  &\nodata & 14:31:08.10 & +05:11:21.18 & \nodata & 216.42 & 226.17 & \nodata  & \nodata & \nodata \\ 
b  &\nodata & 14:31:10.86 & +05:11:15.80 & \nodata & 158.20 & 177.62 &  \nodata  & \nodata  & \nodata  \\ 
J1500+0031 & 23 April 2003 & 15:00:55.34 & +00:31:58.52 & N & 141.31 & 145.90 & $>$20.21 & 2.0 & $>$3.3 \\ 
J1507+6003 & 04 June 2004 &  15:07:44.31 & +60:03:12.68 & N & 181.08 & 189.23 & $>$20.18  & 0.9 & $>$3.3 \\ 
J1527+4352 & 02 June 2004 &  15:27:51.49 & +43:52:4.79 & N &138.95 & 140.11 & $>$20.35  & 0.9 & $>$3.6 \\ 
J1554+3942 & 03 June 2004 &  \nodata  & \nodata  & Y & \nodata   & \nodata  & $>$20.23 & 0.9 & $>$3.3   \\ 
a  &\nodata & 15:54:17.02 & +39:42:27.44 & \nodata & 35.95 & 55.94 & \nodata  &\nodata &\nodata   \\ 
b  &\nodata & 15:54:17.81 & +39:42:18.95 & \nodata & 98.92 & 113.29 & \nodata  & \nodata &\nodata  \\ 
J1557+4657 & 04 June 2004 &  15:57:24.61 & +46:57:54.29 & N & 198.10 & 204.20 & $>$20.14  & 0.9 & $>$3.2 \\ 
J1618+5210 & 04 June 2004 &  16:18:55.60 & +52:10:41.40 & N & 108.26 & 112.29 & $>$20.21  & 0.9 & $>$3.3 \\ 
J1641+4209 & 22 April 2003 & 16:41:30.14 & +42:09:25.99 & N & 260.20 & 263.68 & $>$20.37  & 1.0 & $>$3.6 \\ 
J1648+3623 & 04 June 2004 &\nodata  &\nodata & Y & \nodata &\nodata & $>$20.12 & 0.9 & $>$3.2 \\
a  &\nodata & 16:48:51.58 &  +36:23:39.10 & \nodata & 121.33 & 128.45 &\nodata &\nodata &\nodata \\
b  &\nodata & 16:48:53.02 & +36:23:24.56 & \nodata & 123.58 & 128.96 &\nodata &\nodata &\nodata \\
J1700+3830 & 20 April 2003 & 17:00:19.95 & +38:30:33.93 & N & 428.65 & 430.20 & $>$20.40 & 1.2 & $>$3.6 \\ 
J1711+3047 & 21 April 2003 & 17:11:26.65 & +30:47:45.89 & N & 124.27 & 124.77 & $>$20.41  & 1.2 & $>$3.7 \\ 
J2221$-$0901 & 02 June 2004 &  22:21:48.04 & $-$09:01:58.95 & N & 224.64 & 234.50 &$>$20.32  & 0.9 & $>$3.5 \\ 
J2242$-$0808$\tablenotemark{b}$ & 01 \& 04 June 2004 &  22:42:34.05 & $-$08:08:21.86 & N & 125.64 & 129.97 & $>$20.63 & 0.9 & $>$4.1
\enddata
\tablenotetext{a}{J1350+0352 was observed both 20 April 2003 and 02 June 2004, for a total integration time of 61 minutes with no detection}
\tablenotetext{b}{J2242$-$0808 was observed on two nights in 2004 for a total integration time of 29 minutes}
\label{nondetections}
\end{deluxetable}

\begin{deluxetable}{lcc}
\tablewidth{0pt}
\tablecolumns{4}
\tabletypesize{\small}
\tablecaption{Extinction for $K\,<\,17.5$ Galaxies}
\tablehead{
\colhead{Name} & \colhead{$K$} & \colhead{Extinction}
}
\startdata
0941+0127 &  17.17 & 0.32 \\
0958+0324 &  17.46 & 0.08 \\
1155+0305 &  17.36 & 0.10 \\
1208+4943 &  17.03 & 0.06 \\
1332+0101 &  17.05 & 0.08 \\
1400+0053 &  17.23 & 0.10 \\
1408+0116 &  17.08 & 0.11 \\
1423+0139 &  17.37 & 0.09 \\
1532+4432 &  17.16 & 0.05 \\
1541+5259 &  17.24 & 0.03 \\
1548+0036 &  16.83 & 0.23 \\
1604-0013 &  17.13 & 0.34 \\
1649+3350 &  16.86 & 0.06
\enddata
\label{extinct}
\end{deluxetable}

\begin{deluxetable}{lccc}
\tablewidth{0pt}
\tablecolumns{4}
\tabletypesize{\small}
\tablecaption{Radio Spectral Index}
\tablehead{
\colhead{Name} & \colhead{FIRST $1.4 \, $GHz (mJy)} & \colhead{Texas $365 \,$ MHz (mJy)$\tablenotemark{1}$} & \colhead{$\alpha^{1400}_{365}$}
}
\startdata
J0742+3256 & 144.60 & 426.0 & 0.80 \\
J0831+5210 & 246.00 & 837.0 & 0.91 \\
J0941+0127 & 120.56 & 290.0 & 0.65 \\
J0958+0324 & 638.00 & 2666.6 & 1.06 \\
J1022+0357 & 200.86 & 823.0 & 1.05 \\
J1028+0144 & 114.81 & 391.0 & 0.91 \\
J1044+0538 & 130.20 & \nodata & \nodata \\
J1047+0216 & 125.81 & 311.0 & 0.67 \\
J1102+0250 & 161.00 & 1626.0 & 1.72 \\
J1123+0530 & 1743.00 & 5903.0 & 0.91 \\
J1135+0548 & 212.60 & 982.0 & 1.14 \\
J1144+0254 & 119.18 & 445.0 & 0.98 \\
J1155+0305 & 162.80 & 591.0 & 0.96 \\
J1208+0414 & 641.00 & 2212.0 & 0.92 \\
J1208+4943 & 198.98 & 678.0 & 0.92 \\
J1221+0248 & 142.87 & 510.0 & 0.95 \\
J1234+0024 & 100.55 & 700.0 & 1.44 \\
J1236+0150 & 154.53 & 558.0 & 0.96 \\
J1237+0135 & 426.00 & 1971.0 & 1.14 \\
J1240-0017 & 150.90 & 419.0 & 0.76 \\
J1250+6043 & 304.05 & 725.0 & 0.65 \\
J1259+0559 & 178.90 & 806.0 & 1.12 \\
J1303+0026 & 104.04 & 401.0 & 1.00 \\
J1308-0022 & 241.17 & \nodata & \nodata \\
J1312+0009 & 112.60 & 490.0 & 1.09 \\
J1313+6250 & 132.18 & 476.0 & 0.95 \\
J1314+0330 & 250.05 & 845.0 & 0.91 \\
J1315+0533 & 146.96 & 423.0 & 0.79 \\
J1329+0133 & 102.64 & \nodata & \nodata \\
J1332+0101 & 411.00 & 1430.0 & 0.93 \\
J1336+0207 & 127.76 & 410.0 & 0.87 \\
J1350+0352 & 104.28 & 546.0 & 1.23 \\
J1400+0053 & 130.80 & \nodata & \nodata \\
J1402+0342 & 540.40 & 1192.0 & 0.59 \\
J1403+6048 & 794.62 & 1946.0 & 0.67 \\
J1408+0116 & 612.75 & 1325.0 & 0.57 \\
J1411+0124 & 186.50 & 1029.0 & 1.27 \\
J1421+0248 & 342.63 & 984.0 & 0.78 \\
J1423+0139 & 211.65 & 394.0 & 0.46 \\
J1431+0511 & 226.17 & 956.0 & 1.07 \\
J1438+0150 & 118.50 & 413.0 & 0.93 \\
J1438+6149 & 121.27 & \nodata & \nodata \\
J1451+5404 & 554.63 & 2108.0 & 0.99 \\
J1452+0032 & 639.00 & 3133.0 & 1.18 \\
J1500+0031 & 145.90 & 549.0 & 0.99 \\
J1507+6003 & 189.23 & \nodata & \nodata \\
J1510+5244 & 505.76 & 1596.0 & 0.85 \\
J1515+5744 & 144.16 & 400.0 & 0.76 \\
J1523-0018 & 231.44 & 444.0 & 0.48 \\
J1526+0408 & 156.40 & 504.0 & 0.87 \\
J1527+4352 & 140.11 & \nodata & \nodata \\
J1532+4432 & 249.65 & \nodata & \nodata \\
J1541+5259 & 193.00 & 565.0 & 0.80 \\
J1543+5711 & 103.39 & 279.0 & 0.74 \\
J1547+4839 & 214.75 & \nodata & \nodata \\
J1548+0036 & 126.00 & 313.0 & 0.68 \\
J1548-0033 & 432.00 & 1221.0 & 0.77 \\
J1549+4719 & 106.33 & 252.0 & 0.64 \\
J1554+4729 & 149.73 & 612.0 & 1.05 \\
J1554+3942 & 169.23 & 697.0 & 1.05 \\
J1557+4657 & 204.20 & 756.0 & 0.97 \\
J1559+5011 & 130.28 & 419.0 & 0.87 \\
J1604+4746 & 366.10 & \nodata & \nodata \\
J1604-0013 & 128.28 & \nodata & \nodata \\
J1606+4751 & 109.02 & 670.0 & 1.35 \\
J1609+3700 & 102.52 & 206.0 & 0.52 \\
J1617+4848 & 244.32 & \nodata & \nodata \\
J1618+5210 & 112.29 & 294.0 & 0.72 \\
J1629+4937 & 197.05 & 499.0 & 0.69 \\
J1632+4056 & 203.15 & \nodata & \nodata \\
J1634+4155 & 263.47 & 1155.0 & 1.10 \\
J1636+4808 & 263.30 & 981.0 & 0.98 \\
J1637+3223 & 147.47 & 438.0 & 0.81 \\
J1641+4209 & 263.68 & 1186.0 & 1.12 \\
J1643+4518 & 110.39 & 304.0 & 0.75 \\
J1645+4152 & 115.96 & \nodata & \nodata \\
J1648+4233 & 169.30 & \nodata & \nodata \\
J1648+3623 & 257.00 & 1197.0 & 1.14 \\
J1649+3350 & 163.91 & \nodata & \nodata \\
J1654+4125 & 228.56 & 600.0 & 0.72 \\
J1655+2723 & 169.53 & \nodata & \nodata \\
J1656+2707 & 164.76 & 347.0 & 0.55 \\
J1700+3830 & 430.20 & 868.0 & 0.52 \\
J1707+2408 & 169.15 & 787.0 & 1.14 \\
J1711+3047 & 124.77 & 407.0 & 0.88 \\
J1715+3027 & 385.34 & 1108.0 & 0.79 \\
J2059-0603 & 161.72 & \nodata & \nodata \\
J2107-0701 & 550.60 & 2373.0 & 1.09 \\
J2221-0901 & 234.50 & 930.0 & 1.02 \\
J2223-0757 & 108.34 & 361.0 & 0.90 \\
J2242-0808 & 129.97 & 573.0 & 1.10 \\
J2247-0910 & 103.37 & \nodata & \nodata \\
J2309-0846 & 111.67 & \nodata & \nodata \\
J2316-0846 & 105.40 & 338.0 & 0.87 \\
J2336-0838 & 244.82 & 1137.0 & 1.14 \\
J2337-0852 & 120.14 & 507.0 & 1.07
\enddata
\tablenotetext{1}{From \citet[]{Dou:96}}
\label{specindex}
\end{deluxetable}
\begin{deluxetable}{lcccccccc}
\tablewidth{0pt}
\tablecolumns{9}
\tabletypesize{\tiny}
\tablecaption{Radio Spectral Index}
\tablehead{
\colhead{Name} & \colhead{$4.85 \, $GHz (mJy)$\tablenotemark{1}$} &\colhead{FIRST $1.4 \, $GHz (mJy)} & \colhead{Texas $365 \, $MHz (mJy)} &  \colhead{$151 \, $MHz (mJy)} & \colhead{$\alpha^{1400}_{365}$} & \colhead{$\alpha^{4850}_{1400}$} & \colhead{$\alpha^{1400}_{151}$} & \colhead{$\alpha^{365}_{151}$}
}
\startdata
J1123+0530 & 526 $\pm$ 73 & 1743.0 $\pm$ 69.7 & 5903 $\pm$ 140 & \nodata & 0.9 $\pm$ 0.041 & 1.0 $\pm$ 0.12 & \nodata & \nodata \\
J1208+4943 & 58 $\pm$ 8 & 199.0 $\pm$ 8.0 & 678 $\pm$ 34 & 990 $\pm$ 103 $\tablenotemark{2}$     & 0.9 $\pm$ 0.053 & 1.0 $\pm$ 0.12 & 0.7 $\pm$ 0.052 & 0.4 $\pm$ 0.13 \\
J1313+6250 & 29 $\pm$ 6 & 132.2 $\pm$ 5.3 & 476 $\pm$ 19 & 970 $\pm$ 48.5$\tablenotemark{3}$ & 1.0 $\pm$ 0.048 & 1.2 $\pm$ 0.17 & 0.9 $\pm$ 0.032 & 0.8 $\pm$ 0.072 \\
J1403+6048 & 268 $\pm$ 26 & 794.6 $\pm$ 31.8 & 1946 $\pm$ 31 & 3070 $\pm$ 154$\tablenotemark{3}$ & 0.7 $\pm$ 0.039 & 0.9 $\pm$ 0.088 & 0.6 $\pm$ 0.032 & 0.5 $\pm$ 0.060 \\
J1543+5711 & 25 $\pm$ 5 & 103.4 $\pm$ 4.1 & 279 $\pm$ 15 & 490 $\pm$ 24.5$\tablenotemark{3}$ & 0.7 $\pm$ 0.055 & 1.1 $\pm$ 0.17 & 0.7 $\pm$ 0.032 & 0.6 $\pm$ 0.083 \\
J1554+472 & 34 $\pm$ 6 & 149.7 $\pm$ 6.0 & 612 $\pm$ 18 & 1260 $\pm$ 126$\tablenotemark{2}$ & 1.0 $\pm$ 0.043 & 1.2 $\pm$ 0.15 & 1.0 $\pm$ 0.050 & 0.8 $\pm$ 0.12 \\
J1557+4657 & 39 $\pm$ 6 & 204.2 $\pm$ 8.2 & 756 $\pm$ 19 & 1580 $\pm$ 158$\tablenotemark{2}$ & 1.0 $\pm$ 0.042 & 1.3 $\pm$ 0.13 & 0.9 $\pm$ 0.050 & 0.8 $\pm$ 0.12 \\
J1559+5011 & 37 $\pm$ 6 & 130.3 $\pm$ 5.2 & 419 $\pm$ 19 & 960 $\pm$ 103$\tablenotemark{2}$ & 0.9 $\pm$ 0.050 & 1.0 $\pm$ 0.14 & 0.9 $\pm$ 0.053 & 0.9 $\pm$ 0.13 \\
J1606+4751 & 54 $\pm$ 8 & 109.0 $\pm$ 4.4 & 670 $\pm$ 44 & 1660 $\pm$ 166$\tablenotemark{2}$ & 1.4 $\pm$ 0.061 & 0.6 $\pm$ 0.13 & 1.2 $\pm$ 0.050 & 1.0 $\pm$ 0.14 \\
J1618+5210 & 45 $\pm$ 7 & 112.3 $\pm$ 4.5 & 294.0 $\pm$ 20 & 590 $\pm$ 29.5$\tablenotemark{3}$ & 0.7 $\pm$ 0.063 & 0.7 $\pm$ 0.13 & 0.7 $\pm$ 0.032 & 0.8 $\pm$ 0.096 \\
J1629+4937 & 87 $\pm$ 11 & 197.0 $\pm$ 7.9 & 499 $\pm$ 24 & 770 $\pm$ 103$\tablenotemark{2}$ & 0.7 $\pm$ 0.052 & 0.7 $\pm$ 0.11 & 0.6 $\pm$ 0.064 & 0.5 $\pm$ 0.16 \\
J1637+3223 & 49 $\pm$ 8 & 147.5 $\pm$ 5.9 & 438 $\pm$ 24 & 660 $\pm$ 103$\tablenotemark{2}$ & 0.8 $\pm$ 0.055 & 0.9 $\pm$ 0.14 & 0.7 $\pm$ 0.074 & 0.5 $\pm$ 0.19 \\
J1641+4209 & 67 $\pm$ 9 & 263.7 $\pm$ 10.5 & 1186 $\pm$ 25 & 1810 $\pm$ 181$\tablenotemark{2}$ & 1.1 $\pm$ 0.040 & 1.1 $\pm$ 0.12 & 0.9 $\pm$ 0.050 & 0.5 $\pm$ 0.12 \\
J1643+4518 & 40 $\pm$ 7 & 110.4 $\pm$ 4.4 & 304 $\pm$ 20 & 480 $\pm$ 103$\tablenotemark{2}$ & 0.8 $\pm$ 0.061 & 0.8 $\pm$ 0.15 & 0.7 $\pm$ 0.099 & 0.5 $\pm$ 0.25 \\
J1711+3047 & 41 $\pm$ 7 & 124.8 $\pm$ 5.0 & 407 $\pm$ 35 & 670 $\pm$ 103$\tablenotemark{2}$ & 0.9 $\pm$ 0.074 & 0.9 $\pm$ 0.14 & 0.8 $\pm$ 0.073 & 0.6 $\pm$ 0.20 
\enddata
\tablenotetext{1}{From \citet[]{Gre:91}}
\tablenotetext{2}{From \citet[]{Hal:88}}
\tablenotetext{3}{From \citet[]{Hal:90}}
\label{bigspecindex}
\end{deluxetable}
\begin{deluxetable}{ccc}
\tablewidth{0pt}
\tablecolumns{3}
\tabletypesize{\small}
\tablecaption{Differential $K$-band Number Counts}
\tablehead{
\colhead{K} & \colhead{Galaxies} & \colhead{Stars}
}
\startdata
11.625 & 1 & 42 \\
11.875 & 1 & 54 \\
12.125 & 4 & 66 \\
12.375 & 4 & 77 \\
12.625 & 8 & 87 \\
12.875 & 8 & 119 \\
13.125 & 11 & 155 \\
13.375 & 22 & 167 \\
13.625 & 28 & 184 \\
13.875 & 32 & 179 \\
14.125 & 43 & 235 \\
14.375 & 62 & 259 \\
14.625 & 85 & 339 \\
14.875 & 117 & 345 \\
15.125 & 199 & 397 \\
15.375 & 257 & 477 \\
15.625 & 381 & 494 \\
15.875 & 518 & 531 \\
16.125 & 655 & 545 \\
16.375 & 858 & 693 \\
16.625 & 1223 & 713 \\
16.875 & 1490 & 751 \\
17.125 & 2204 & 597 \\
17.375 & 2786 & 496 \\
17.625 & 3256 & 475 \\
17.875 & 3784 & 443 \\
18.125 & 4230 & 376 \\
18.375 & 4402 & 274 \\
18.625 & 4386 & 237 \\
18.875 & 3813 & 145 \\
19.125 & 2820 & 107 \\
19.375 & 1026 & 31 \\
19.625 & 210 & 10 \\
19.875 & 191 & 3 \\
20.125 & 113 & 2 \\
20.375 & 110 & 0 \\
20.625 & 87 & 0 \\
20.875 & 73 & 1 
\enddata
\label{numcounts}
\end{deluxetable}

\end{document}